\documentclass[superscriptaddress,groupedaddress,nofootnoteinbib,11pt]{article}
\pdfoutput=1 
\usepackage{jheppub}

\usepackage{mathtools,slashed,mathrsfs}
\usepackage[caption=false]{subfig}
\usepackage{dcolumn}
\usepackage{multirow}
\usepackage{tabularx}
\usepackage{booktabs}
\usepackage{bm}
\usepackage{comment}
\usepackage{setspace}
\usepackage[dvipsnames]{xcolor}
\usepackage[normalem]{ulem} 
\usepackage{enumerate}
\usepackage{siunitx}

\definecolor{mypurple}{RGB}{164,64,214}

\newcommand{\ffs}{f_\mathrm{FS}}

\newcommand\nn{\nonumber}
\newcommand{\bea}{\begin{eqnarray}}
\newcommand{\eea}{ \end{eqnarray} }

\def\d{\partial}
\def\l{\left(}
\def\r{\right)}

\begin{document}

\title{Precision Early Universe Cosmology from Stochastic Gravitational Waves}

\author[a]{Dawid Brzeminski,}
\author[a]{Anson Hook,}
\author[a]{and Gustavo Marques-Tavares}

\affiliation[a]{Maryland Center for Fundamental Physics, University of Maryland, College Park, MD 20742}
\emailAdd{dbrzemin@umd.edu}
\emailAdd{hook@umd.edu}
\emailAdd{gusmt@umd.edu}

\abstract{
The causal tail of stochastic gravitational waves can be used to probe the energy density in free streaming relativistic species as well as measure $g_\star(T)$ and beta functions $\beta(T)$ as a function of temperature.
In the event of the discovery of loud stochastic gravitational waves, we demonstrate that LISA can measure the free streaming fraction of the universe down to the the $10^{-3}$ level, 100 times more sensitive than current constraints.  Additionally, it would be sensitive to $\mathcal{O}(1)$  deviations of $g_\star$ and the QCD $\beta$ function from their Standard Model value at temperatures $\sim 10^5$ GeV.
In this case, many motivated models such as split SUSY and other solutions to the Electroweak Hierarchy problem would be tested.
Future detectors, such as DECIGO, would be 100 times more sensitive than LISA to these effects and be capable of testing other motivated scenarios such as WIMPs and axions.
The amazing prospect of using precision gravitational wave measurements to test such well motivated theories provides a benchmark to aim for when developing a precise understanding of the gravitational wave spectrum both experimentally and theoretically.
}

\maketitle

\section{Introduction}

With the discovery of gravitational waves by LIGO, we have added a new window with which to view the universe~\cite{LIGOScientific:2016aoc}.  Due to its extremely weak coupling, gravitational waves (GWs) provide an almost unperturbed view of the universe.  As such, they represent a unique opportunity with which to learn about early universe cosmology.  Additionally, many new gravitational wave detectors are expected to be built in the future, such as LISA~\cite{LISA:2017pwj}, BBO~\cite{Harry:2006fi}, MAGIS~\cite{Graham:2017pmn} and DECIGO~\cite{Kawamura:2011zz}, which will enhance our ability to observe GWs by many orders of magnitude.

If observed, stochastic gravitational waves would be the gravitational equivalent of the Cosmic Microwave Background (CMB), and thus, much could be learned from their detection.  Similar to the CMB, the first thing we would likely learn is the mechanism by which GWs were generated.  In the case of the CMB, it was black body radiation, while for GWs there exist a vast number of possibilities for production mechanisms~\cite{Grishchuk:1974ny,Starobinsky:1979ty,Rubakov:1982df,Guzzetti:2016mkm,Khlebnikov:1997di,Easther:2006gt,Easther:2006vd,GarciaBellido:2007dg,GarciaBellido:2007af,Dufaux:2007pt,Caprini:2019egz,Caprini:2018mtu,Christensen:2018iqi,Acquaviva:2002ud,Mollerach:2003nq,Baumann:2007zm,Espinosa:2018eve,Kohri:2018awv,Domenech:2019quo,Hindmarsh:2020hop,Dunsky:2021tih}.  Aside from having many different possible sources, the theories that give rise to stochastic GWs are also well motivated~\cite{Grojean:2006bp,Schwaller:2015tja,Chang:2019mza,Gouttenoire:2019rtn,Cui:2019kkd,Buchmuller:2019gfy,Dror:2019syi,Dunsky:2019upk,Blasi:2020wpy,Machado:2019xuc}.
As with the CMB, angular anisotropies would encode information about primordial fluctuations and could be used to learn more about inflation~\cite{Geller:2018mwu}.
For example, the lack of a correlation between stochastic GWs and the CMB could indicate the presence of an additional light particle present during inflation.

Famously, propagation effects also leave an imprint on a cosmic background. 
For the CMB, this allowed a precise determination of the dark energy density and also gives sensitive to the matter power spectrum at lower redshift through lensing effects. In addition, future probes of the 21 cm absorption line can be used to determine other properties of the early universe.
In the case of stochastic GWs, there are two main propagation effects that can change the GW spectrum coming from the equation of state of the universe ($w(T)$) and from relativistic free-streaming particles ($f_{FS} (T) = \rho_{FS}/\rho_\text{total}$).  Both $w(T)$ and $f_{FS}(T)$, as well as changes to them, can be in principle observed from the frequency spectrum of a stochastic GW background~\cite{Hook:2020phx,Seto:2003kc,Boyle:2005se,Watanabe:2006qe,Boyle:2007zx,Jinno:2012xb,Caprini:2015zlo,Geller:2018mwu,Saikawa:2018rcs,Cui:2018rwi,Caldwell:2018giq,DEramo:2019tit,Figueroa:2019paj,Auclair:2019wcv,Chang:2019mza,Caprini:2019egz,Gouttenoire:2019kij,Gouttenoire:2019rtn,Blasi:2020wpy,Domenech:2020kqm,Caprini:2018mtu,Allahverdi:2020bys,Cai:2019cdl}. 

The expected sensitivity of 21 cm experiments is due to knowing that the CMB follows a blackbody spectrum ahead of time, so that deviations from this predicted spectrum can be attributed to propagation effects.
In contrast, the stochastic GW spectrum is completely unknown.  Even if observed, it will be a highly non-trivial task to make a postdiction for what different parts of the spectrum should look like in the absence of propagation effects.  As such, it would appear that propagation effects would be impossible to disentangle from the source generating the spectrum.  However, there is an important case where we do know the shape of the spectrum ahead of time, the low frequency limit of causally produced GWs~\cite{Caprini:2009fx,Hook:2020phx,Cai:2019cdl}.

If the source producing GWs is only active for a short time period, this source is necessarily uncorrelated on distance scales longer than the time it took to produce the GWs (e.g. distances longer than the Hubble size are necessarily uncorrelated).  This fact determines the shape of the GW spectrum at large wavelengths completely independent of the details of how the GW was generated.  
As an example, the low frequency tail of GWs generated by phase transitions is universal regardless of the details of the phase transition itself.  In contrast, the low frequency tail of inflationary GWs is model dependent and cannot be predicted a priori.
Thus, by studying the low frequency tail of stochastic GWs generated by a time localized source, one can learn about early universe dynamics through propagation effects on the GW spectrum.

In this paper, we study how sensitive future experiments, such as LISA and DECIGO, will be to propagation effects and provide well-motivated benchmark models to illustrate what can be achieved with varying levels of sensitivity.  We find that in the event of a large stochastic GW background, LISA could measure $w(T)$ and $f_{FS}$ with an accuracy of $\sim 10^{-3}$ over a temperature range of $10^4 - 10^6$ GeV. This demonstrates that LISA's sensitivity to $f_{FS}$ can be $10-100$ times more sensitive than even CMB S4 experiments.
Future experiments such as DECIGO can measure $w(T)$ and $f_{FS}$ with an accuracy of $10^{-6}$ and extended sensitivity to an even higher temperature range of $10^4 - 10^8$ GeV.

In the case of a radiation dominated universe, being able to measure $w(T)$ at the $10^{-4} - 10^{-3}$ level is a complete game changer.  While one might think that $w = 1/3$ at high temperatures, due to the scale anomaly $w(T) - 1/3 \sim \beta(T)/g_\star(T)$ where $\beta$ is a combination of Standard Model (SM) beta functions with the QCD beta function giving the largest contribution.  For the SM, $w(T)-1/3 \sim 5 \times 10^{-4}$.  Thus LISA could, in principle, test the SM prediction that $w \ne 1/3$ in a radiation dominated universe.

Aside from the possible spectacular confirmation of a SM prediction, these measurements can test some of the most well motivated particle physics models related to the Electroweak Hierarchy problem~\cite{Sundrum:2005jf,Csaki:2018muy} and dark matter~\cite{Lin:2019uvt}.  Dimensional analysis predicts that the physics responsible for the mass of the Higgs boson should be at the TeV scale.  Solutions to this problem often require doubling the number of particles in the SM.  At temperatures above the TeV scale, this would double $g_\star$ and/or $\beta$ and hence change the predicted value of $w(T > \text{TeV})$ by $\mathcal{O}(10^{-4} - 10^{-3})$, something testable by LISA.

There is also strong reason to expect the mass of dark matter to be below the 100 TeV scale.  If what is responsible for dark matter was ever in thermal contact with the SM, then modulo some mild assumptions, unitarity requires that the energy scale of dark matter be below 100 TeV~\cite{Griest:1989wd}.  A classic motivated example of such a dark matter particle is the famous weakly interacting massive particle (WIMP).  WIMPs have a TeV scale mass whose precise value depends on their representation under $SU(2)$.  In the early universe, at temperatures $T \gtrsim$ 100 TeV, dark matter would be in thermal equilibrium and change $g_\star$ by $\sim 1$.  WIMP dark matter would thus be guaranteed to change $w(T)$ by $\mathcal{O}(10^{-5})$.  While LISA cannot reach this sensitivity, future detectors such as DECIGO could potentially test all thermal dark matter candidates.

Precision measurements of $f_{FS}$ in the early universe are equally important.  
It is worth prefacing the discussion of $f_{FS}$ by noting that we have no direct observations of the radiation domination epoch with $f_{FS} < 0.3$.  Thus even a negative measurement would indicate a phase of the universe that we have never observationally seen before.
One important benchmark is $f_{FS} \sim 10^{-2}$, which is the value of a single new degree of freedom that was in thermal equilibrium but has since left equilibrium.  In many motivated models, such as the QCD axion~\cite{Peccei:1977hh,Peccei:1977ur,Weinberg:1977ma,Wilczek:1977pj}, there is at least one new particle of this sort and reaching sensitivities of the order of $\mathcal{O}(10^{-2})$ would constitute an important test of these theories.
Additionally, as GWs are themselves free streaming, one could even measure the backreaction of GWs on themselves if the stochastic GW signal was loud enough.
Like the neutrinos, new free streaming particles beyond the Standard Model may become non-relativistic after going out of equilibrium with the SM bath.  We also explore how this behavior changes the effect of free streaming particles on GWs.

In Sec.~\ref{Sec: sensitivity}, we discuss the sensitivity of GW detectors to $w(T)$ and $f_{FS}(T)$.
In Sec.~\ref{Sec: application}, we discuss the impact of these measurements on well motivated models.
In Sec.~\ref{Sec: free}, we discuss in more detail how massive free streaming particles change the GW spectrum.
We finish with concluding remarks in Sec.~\ref{Sec: conclusion}

\section{Sensitivity of future experiments to the causal tail of stochastic GWs}
\label{Sec: sensitivity}

In this section we discuss the sensitivity of GW experiments to the low frequency tail of causally produced GWs.  It is well known that the low frequency tail of any time localized, causally produced GW has a fixed form regardless of how it was produced~\cite{Caprini:2009fx}.
The fixed form of this low frequency tail depends on details such as the equation of state $w(T)$ in the era following the production of the GW~\cite{Cai:2019cdl,Hook:2020phx}.  Different values of $w(T)$ give the function form for the gravitational wave power spectrum
\bea
\label{Eq: w-slope}
\Omega_{GW}(k) \propto k^{3 - 2 \l \frac{1 - 3 w}{1 + 3 w} \r } \, , 
\eea
where $w$ is the equation of state when the mode with wave-number $k$ entered the horizon. Aside from the equation of state, the low frequency tail is also sensitive to the free-streaming fraction 
\begin{equation} \label{eq:ffs-definition}
	f_{FS}(T) = \frac{\rho_{FS}(T)}{\rho_\text{total}(T)} \, .
\end{equation}
Weinberg first showed that free streaming particles can dampen inflationary gravitational waves~\cite{Weinberg:2003ur} and similar effects were shown to be present for non-inflationary GWs~\cite{Hook:2020phx}.   As a GW travels, it imparts a quadropole moment on any ambient free streaming particles that subsequently backreacts and suppresses the GW in much the same way that a dielectric suppresses electric fields.  
Assuming radiation domination and relativistic free-streaming particles, the dependence is of the form
\bea
\label{Eq: ffs-slope}
\Omega_{GW}(k) \propto k^{3  +  \frac{16}{5} f_{FS}} \,
\eea
in the limit where $\ffs \ll 1$~\cite{Hook:2020phx}. The explicit form for arbitrary $\ffs$ , including the case where $\ffs \gtrsim 5/32$ when $\Omega_{GW}$ develops an oscillatory pattern, must be calculated numerically (see Eq.~\ref{eq:large-ffs} for its approximate analytical form). We can see from Eq.~\ref{Eq: w-slope} the famous fact that in a radiation dominated era ($w=1/3$) with no relativistic free streaming particles, the spectrum is predicted to scale exactly as $k^3$. Any deviation from $k^3$ scaling would be evidence of free streaming particles or that $w \ne 1/3$.

A transfer function can be used to take into account the effects of $w$ or $f_{FS}$ on the GW spectrum.  Given a GW spectrum that was calculated/numerically simulated in a radiation dominated universe ($\Omega^0_{GW}(k)$), we can take into account $w \ne 1/3$ and/or $f_{FS} \ne 0$  using
\bea \label{Eq: decompose}
\Omega_{GW}(k,\theta) = \Omega^0_{GW}(c(\theta)\, k) \, F(c(\theta) \, k,\theta) \, ,
\eea
where $\theta = w$ or $f_{FS}$, and $c(\theta)$ is a re-scaling of frequencies that arises due to the change in the redshift at which the GWs were produced (since the temperature will evolve differently as a function of redshift for different $w(T)$)\footnote{This is equivalent to a rescaling of the frequency and does not affect the shape of the spectrum.  As the shape of the spectrum will be the main focus of our analysis, and because a shift in the frequency scales is also achieved by rescaling the unknown temperature of the PT, we will neglect this rescaling of frequency from now on.}.
$\Omega^0_{GW}(k)$ is the spectrum for $w=1/3$ and $\ffs = 0$, while $F(k,\theta)$ is a transfer function that encapsulates how the spectrum depends on these additional parameters and will be specified in their respective subsections. 
When estimating the sensitivity of GW detectors to $\theta$, we need to specify $\Omega^0_{GW}(k)$.  

We will take $\Omega^0_{GW}(k)$ to be the gravitational wave spectrum coming from the sound waves of a phase transition, but the results of this paper will apply for any causal source of gravitational waves.  For simplicity, we will use an analytical approximation of numerical results~\cite{Hindmarsh:2017gnf}\footnote{
The dependence of the spectral shape of the sub-horizon modes on the physics of the PT is still an active topic of investigation, for some recent discussion see e.g.~\cite{Ellis:2018mja,Caprini:2019egz,Ellis:2020awk}
}
\bea
\label{Eq: sound}
h^2 \Omega^0_{GW} &=& 1.19 \times 10^{-6} \l \frac{H_{PT}}{\beta} \r \l \frac{\kappa_{v}\alpha}{1 + \alpha} \r^2 \l \frac{100}{g_\star} \r^{1/3} \l \frac{f}{f_\star} \r^3 \l \frac{7}{4 + 3 \l f/f_\star \r^2} \r^{7/2}  , \\
f_\star &=& 8.9 \times 10^{-3} \, \text{mHz} \frac{1}{v_{w}} \l \frac{\beta}{H_{PT}} \r  \l \frac{T_{PT}}{100 \, \text{GeV}} \r \l \frac{g_\star}{100} \r^{1/6} ,
\eea
where $f_\star$ is the peak frequency of the spectrum, $T_{PT}$ ($H_{PT}$) is the temperature (Hubble scale) at which the phase transition took place, $\alpha = \rho_{PT}/\rho_\text{total}$ is the ratio of the energy released by the PT to the total energy density, the time scale of the transition is $1/\beta$, and $\kappa_v = \alpha /(0.73+0.083\sqrt{\alpha} + \alpha )$ is the fraction of the latent heat which gets converted into the bulk motion of the fluid.   In the last expression, we are also assuming a large wall velocity $v_w \sim 1$.  

\subsection{SNR and Fisher matrix}

The visibility of a stochastic GW background is typically encapsulated by its signal-to-noise ratio (SNR).  In a detector like LIGO, with two (or more) independent interferometers, a stochastic GW signal is searched for by comparing the power in two different GW detectors, $1$ and $2$.
The SNR is~\cite{Allen:1997ad,Maggiore:1999vm,Kudoh:2005as}
\bea
\text{SNR}^2 = 2 T \int\limits_{-\infty}^{\infty} \frac{df}{2} \frac{\vert U(f) \vert^{2}}{\vert U(f) \vert^{2} + W(f)} \, ,
\eea
where
\bea
U(f) &=& C_{12}(f) \, , \\
W(f) &=& C_{11}(f) C_{22}(f)+C_{11}(f)N_{2}(f)+C_{22}(f)N_{1}(f)+N_{1}(f)N_{2}(f) .
\eea
The functions $C$ and $N$ are defined as
\bea
C_{IJ}(f;t,t) &=& \int \frac{d\Omega}{4\pi} S_{h}(f) \mathcal{F}_{IJ}(f,\Omega;t,t) \, , \\
\langle\Tilde{n}_{I}^{*}(f)\Tilde{n}_{J}(f')\rangle &=& \frac{1}{2}\delta_{IJ}\delta(f-f')N_{I}(f) \, , \\
\langle\Tilde{h}_{A}^{*}(f,\Omega)\Tilde{h}_{A'}(f',\Omega')\rangle &=& \frac{1}{2} \delta(f-f') \frac{ \delta^{2}(\Omega,\Omega')}{4\pi} \delta_{AA'} S_{h} (\vert f \vert, \Omega) \, ,
\eea
where the indices $I,J$ represent the different detectors used to capture the signal, and $\mathcal{F}_{IJ}$ is the antenna pattern function~\cite{Kudoh:2005as}, which characterizes the overlap in the detectors response to the signal.  $\Tilde{n}$ is the noise while $\Tilde{h}$ is the GW signal. Indices $A,A'$ represent polarization of the GW signal, $f$ is the frequency of the GW while $\Omega$ is the direction it is coming from.
For simplicity we assume that the signal is isotropic so that
\bea \label{eq:OmegaGW}
\Omega_{GW}(f) = \frac{8 \pi G}{3 H_0^2} \frac{d \rho_{GW}}{d \log f} = \frac{4\pi}{3}\frac{f^3 S_{h}(f)}{H_{0}^2}.
\eea
Here $H_0$ is the value of the Hubble parameter today.
As an additional simplification, we also take the various overlap factors between detectors to be identical,  $\mathcal{F}_{12} = \mathcal{F}_{11} = \mathcal{F}_{22}$, and the noise of the two detectors to be the same, $N_{1} = N_{2}$. As a result the SNR simplifies to
\bea
\label{eq: SNR}
\text{SNR}^2 = T \int df \frac{\Omega_{GW}^2}{2 \Omega_{GW}^2 + 2 \Omega_{GW} \Omega_\text{noise} + \Omega_\text{noise}^2},
\eea
where $T$ is the runtime of the experiment and $\Omega_\text{noise}$ is the noise power spectrum (normalized to the critical density as in Eq.~\ref{eq:OmegaGW}). While for future space-base missions this approach is not exactly justified, it recovers the usual SNR expression in the small signal limit~(see e.g.~\cite{Caldwell:2018giq}) (which to our knowledge is the only case that has been widely studied), and at least incorporates some of the important effects of the large signal limit.

We can obtain an intuitive understanding of Eq.~\ref{eq: SNR} by using the fact that the signal is a measurement of the cross correlation between detectors, schematically 
\bea
S = \mu = \langle ( h + n_1) (h + n_2) \rangle \propto \Omega_{GW}.
\eea
Meanwhile, the noise is schematically 
\bea
N^2 &=& \Sigma = \langle ( h + n_1)^2 (h + n_2)^2 \rangle -  \langle ( h + n_1) (h + n_2) \rangle^2 \nn \\
&\propto& 2 \Omega_{GW}^2 + 2 \Omega_{GW} \Omega_\text{noise} + \Omega_\text{noise}^2 ,
\eea
where we have taken the usual approximation that the GW signal is approximately Gaussian.
Taking $SNR = S/N$ for each frequency bin and summing in quadrature over all of the frequency bins gives Eq.~\ref{eq: SNR}.\footnote{
	These expressions will also play a role in forecasting the sensitivity of future space based experiments, even though, as mentioned earlier, they are technically only valid for a LIGO like setup. One can show that they lead to expressions for the fisher matrix that match the literature~\cite{Caldwell:2018giq} both in the small signal limit and in the signal much larger than noise limit, and thus will be used in our analysis.
}

In this article, we will be usually working in the optimistic limit where the signal is large.  As can be seen in Eq.~\ref{eq: SNR}, under this approximation the magnitude of the noise is less important than the runtime and the frequency range under consideration.  Thus, for us, the main difference between experiments when it comes to measuring $w(T)$ or $f_{FS}(T)$ is the difference in frequency ranges that they are sensitive to. 

To estimate the sensitivity of an experiment to $w(T)$ or $f_{FS}(T)$, we will use the Fisher information matrix~(for recent studies forecasting LISA sensitivity to spectral information of GW signals see e.g.~\cite{Caprini:2019pxz,Gowling:2021gcy}), and assume that we will be able to subtract contamination from astrophysical foregrounds . The Fisher matrix is useful because it determines the optimal sensitivity of any unbiased estimator.  Assuming gaussian distributions, the Fisher information matrix can be found from the covariance matrix $\Sigma$ and  mean $\mu$ using
\bea\label{Eq: fisher}
F_{\alpha \beta} &=& \frac{\partial \mu}{\partial \theta_\alpha} \Sigma^{-1} \frac{\partial \mu}{\partial \theta_\alpha} +\frac{1}{2} \text{Tr} \l \Sigma^{-1} \frac{\partial \Sigma}{\partial \theta_\alpha} \Sigma^{-1} \frac{\partial \Sigma}{\partial \theta_\beta} \r \\
&=& T \int df  \frac{\partial \Omega_{GW}}{\partial \theta_\alpha} \frac{\partial \Omega_{GW}}{\partial \theta_\beta}  \frac{4 \Omega_{GW}^2 + 2 \Omega_{GW} \Omega_\text{noise} + \Omega_\text{noise}^2}{\l 2 \Omega_{GW}^2 + 2 \Omega_{GW} \Omega_\text{noise} + \Omega_\text{noise}^2  \r^2},
\eea
where $\theta_\alpha$ are the parameters describing the GW spectrum.  
For example, if we let $w$ be quantity of interest, then we can replace $\theta_{\alpha,\beta} \rightarrow w$ and the Fisher matrix $F = 1/\sigma^2_w$ describes the variance in how well an experiment could measure $w$.
In later subsections, we will utilize Eq.~\ref{Eq: fisher} to calculate the sensitivity of LISA and DECIGO to the equation of state and $f_{FS}$.  
For LISA we take $\Omega_\text{noise}$ from program PTPlot~\cite{Caprini:2019egz} and for DECIGO we take $\Omega_\text{noise}$ from  Ref.~\cite{Kuroyanagi:2010mm}.  In both cases, we take the runtime of the experiment to be $T=10^8$ seconds.

\subsection{Equation of state}

When studying the sensitivity of experiments to the equation of state, we first need to fully specify the GW spectrum, namely we need to determine $F(k,w)$ shown in Eq.~\ref{Eq: decompose} assuming no relativistic free streaming species ($\ffs = 0$).  In the limit that the phase transition occurred very rapidly, $\beta/H_{PT} \gtrsim 1$, and for wavelengths that are larger than the spatial correlations of the source, this function can be analytically shown to be~\cite{Hook:2020phx}
\bea
F(k,\theta=w) = \frac{|j_{\frac{1-3w}{1+3w}} \left ( k \tau_{PT} \right ) |^2 +|y_{\frac{1-3w}{1+3w}} \left ( k \tau_{PT} \right )|^2}{|j_{0} \left ( k \tau_{PT} \right ) |^2 +|y_{0} \left ( k \tau_{PT} \right )|^2} ,
\eea
where $\tau_{PT}$ is the conformal time when the phase transition occurred and $j_n$ and $y_n$ are the spherical Bessel functions.
For simplicity, we will take this to hold even when $\beta/H_{PT} \sim 1$.

With the full GW spectrum in hand, we can use Eq.~\ref{Eq: fisher} to find the sensitivity of any given experiment to the equation of state of the universe.
For simplicity we will express our sensitivity to $\delta w = w- 1/3$ in the form of the variance $\sigma_w$.
In principle, the phase transition parameters ($\alpha$, $\beta$, $T_{PT}$, $\cdots$) can all be determined from the peak of the distribution and thus we will take them to be fixed while studying the sensitivity to $\delta w$.~\footnote{
As discussed earlier, we are also ignoring the shift in frequencies due to the difference in expansion history, which can in principle be reabsorbed by changing the phase transition parameters.}
In order to show how $\sigma_w$ depends on various phase transition parameters, we first fix all parameters except for $T_{PT}$ and a single other parameter and show how $\sigma_w$ varies as a function of a $T_{PT}$ for a few values of the other parameter.  Motivated by one of the louder phase transition models from Ref.~\cite{Caprini:2015zlo}, we will take $\alpha = 1$, $\beta = 3$ and $v_w = 1$ unless otherwise stated.

\begin{figure}[t]
\centering
\includegraphics[width=.7\linewidth]{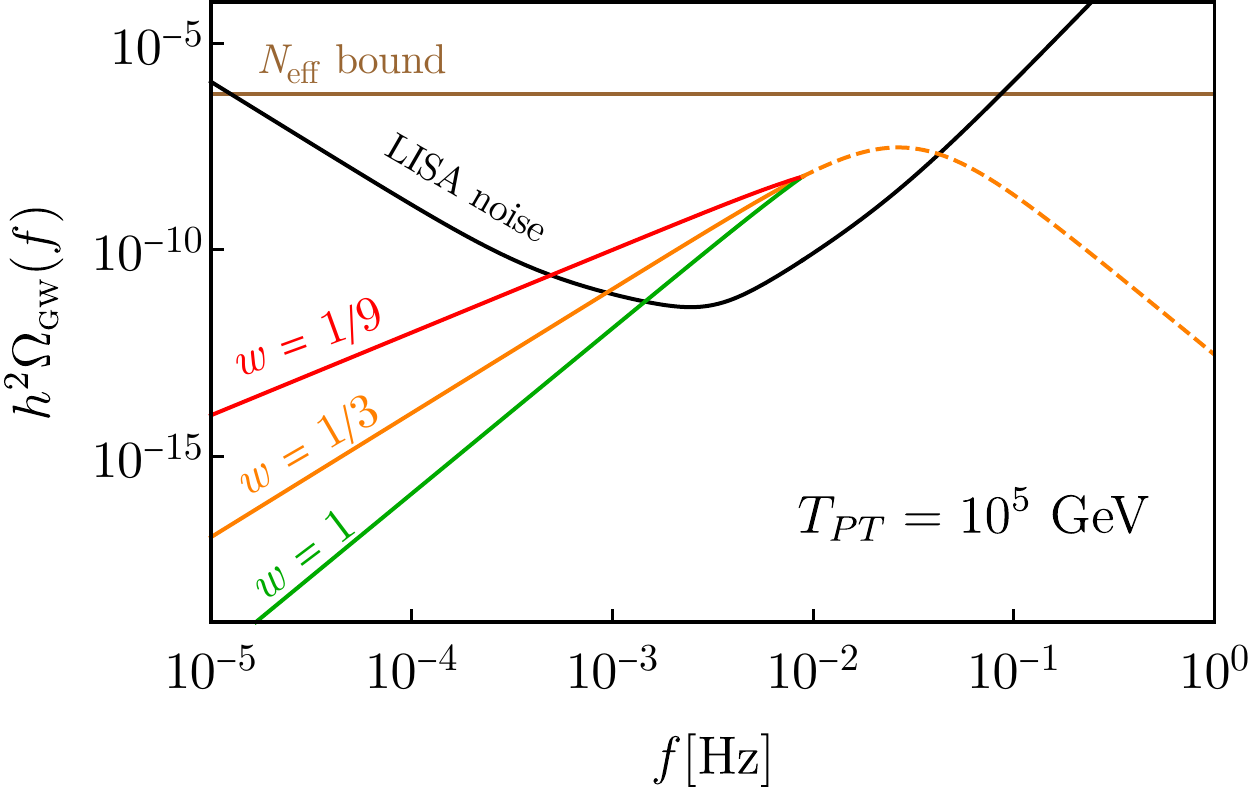}
\caption{A graphical representation of how the gravitational wave spectrum changes for various $w(T)$ compared to the LISA sensitivity curve assuming a runtime of $10^8$ s.  For simplicity we are only showing how the shape changes and neglecting the rescaling of frequencies due to differing expansion histories.  The dashed line indicates modes that were sub-horizon at the time the phase transition took place and their shape is unaffected by the change in expansion history.  The solid lines indicate modes that were super-horizon at the time when the phase transition took place and have their shape distorted by the different expansion history. }
\label{Fig: wspectrum}
\end{figure}

To show visually how $w(T)$ affects the GW spectrum, in Fig.~\ref{Fig: wspectrum} we show how the GW spectrum changes as one changes $w(T)$.
We will be interested in small changes around $w = 1/3$.  Due to the similar scaling of the background noise and the signal, the signal remains above background for a large range of frequencies, leading to sensitivity over a wide range of frequencies (and thus a wide range of temperatures in the early universe).  We will first study the sensitivity $\sigma_w$ to constant deviations in the equation of state before studying the sensitivity to frequency/temperature dependent changes to the equation of state.

\begin{figure}[t]
\centering
\includegraphics[width=.49\linewidth]{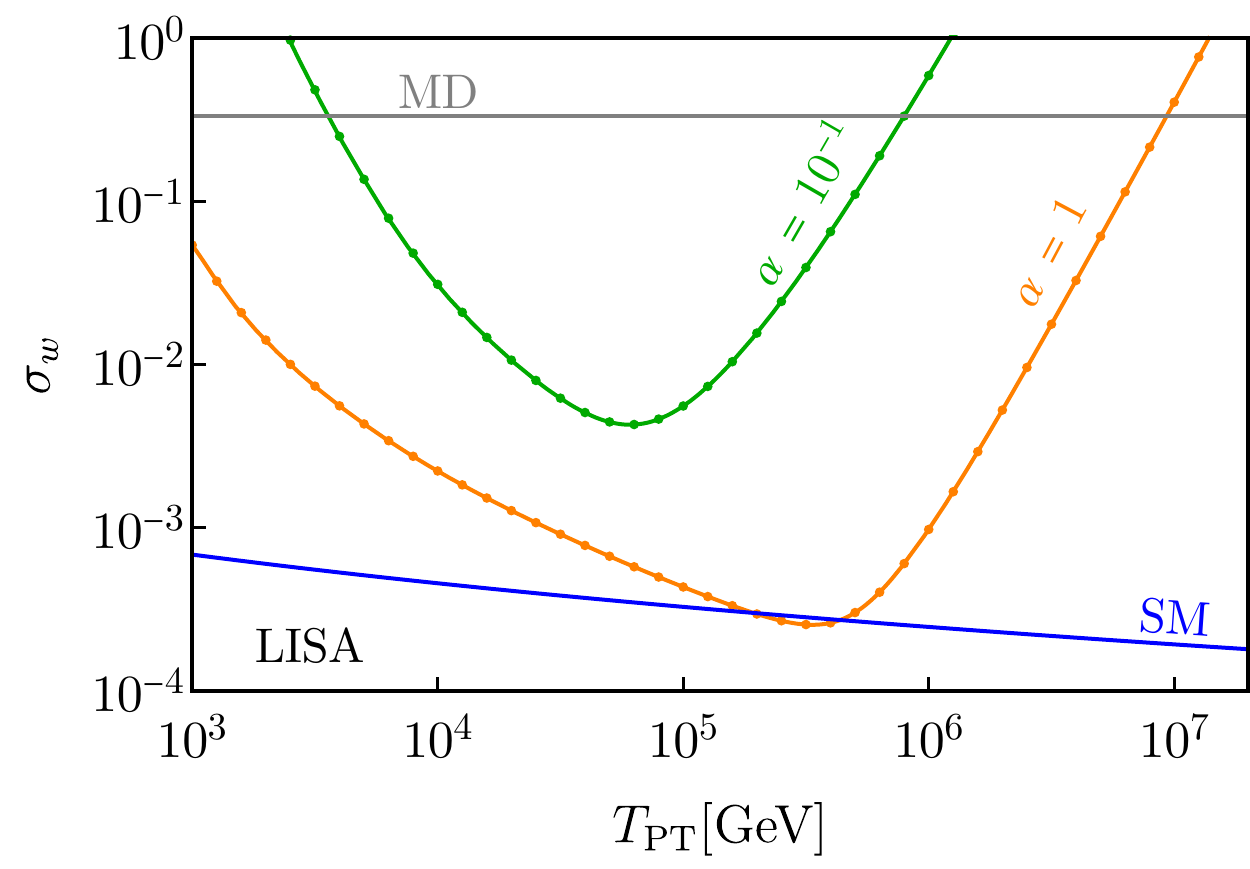}
\includegraphics[width=.49\linewidth]{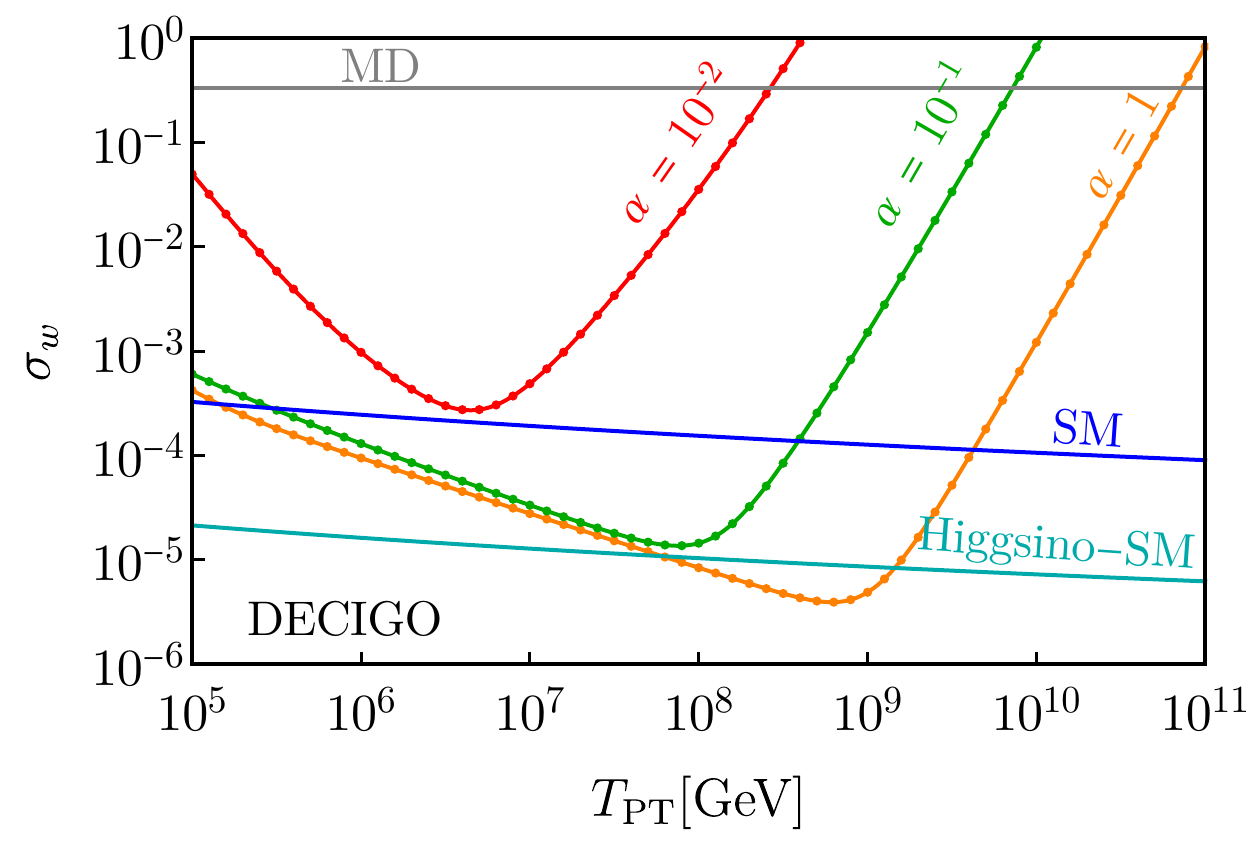}
\caption{The projected sensitivity, calculated using the Fisher matrix, of LISA (left) and DECIGO (right) to a constant equation of state near $w = 1/3$ as a function of the temperature at which the phase transition occurred for various $\alpha$.  $\alpha$ controls the amplitude of a GW.  For reference we show matter domination ($\delta w = -1/3$) as well as the SM and SM plus doublet dark matter predictions for $\delta w$ versus temperature.}
\label{Fig: alphaw}
\end{figure}

We first show how $\sigma_w$ depends on $\alpha$ in Fig.~\ref{Fig: alphaw}.  Varying $\alpha$ changes the amplitude of the signal.  As can be seen from the SNR (Eq.~\ref{eq: SNR}) and the Fisher matrix (Eq.~\ref{Eq: fisher}), the magnitude of the signal is not particularly important if the signal is larger than the background.  This can be seen explicitly in Fig.~\ref{Fig: alphaw} for the DECIGO sensitivity (for LISA the signal is only above the noise for a narrow range of parameters).  For smaller $T_{PT}$, the sensitivity is the same for $\alpha = 1$ and $0.1$.  At higher $T_{PT}$, eventually the signal falls below the background and larger $\alpha$ results in better sensitivity.  For $\alpha = 0.01$, background is important for all temperatures and thus the sensitivity is always worse than larger values of $\alpha$.  

\begin{figure}[t]
\centering
\includegraphics[width=.49\linewidth]{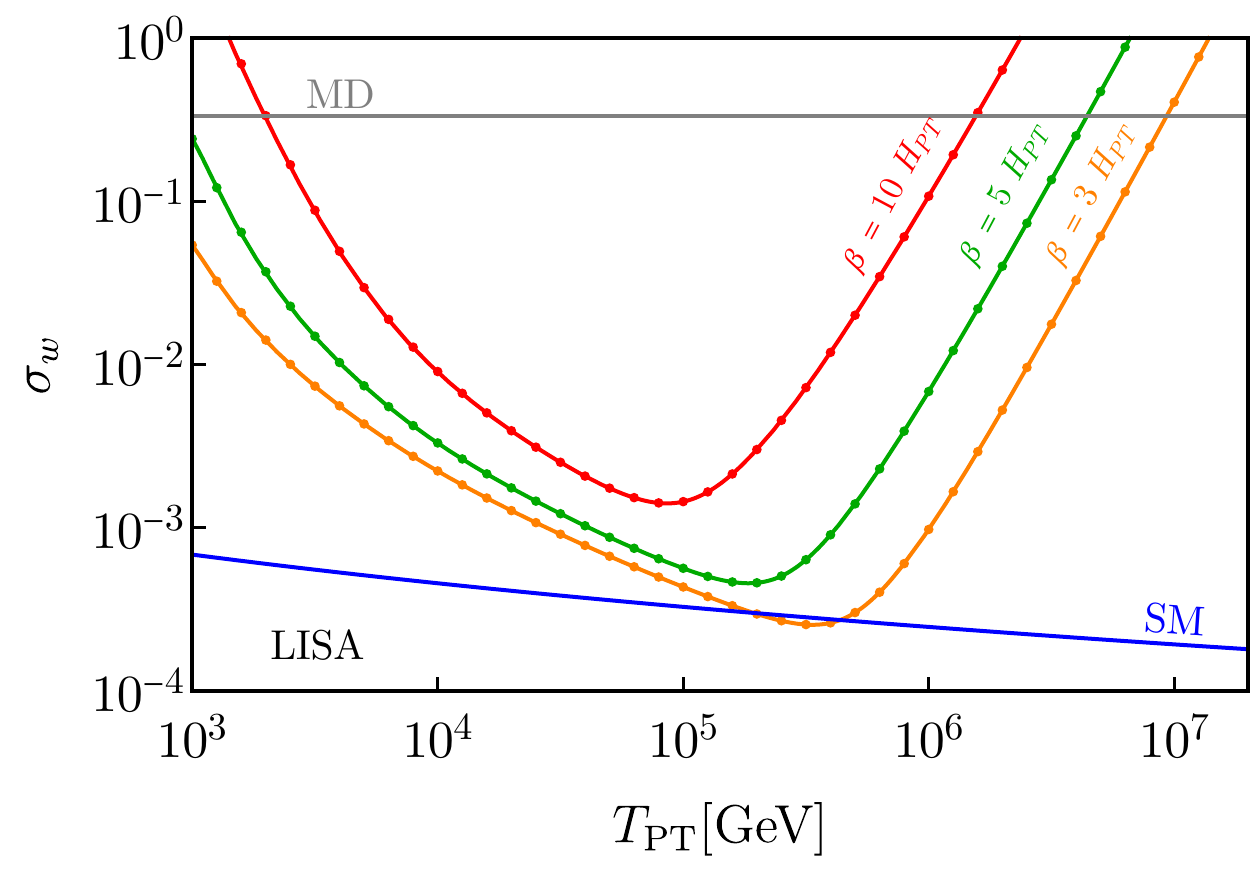}
\includegraphics[width=.49\linewidth]{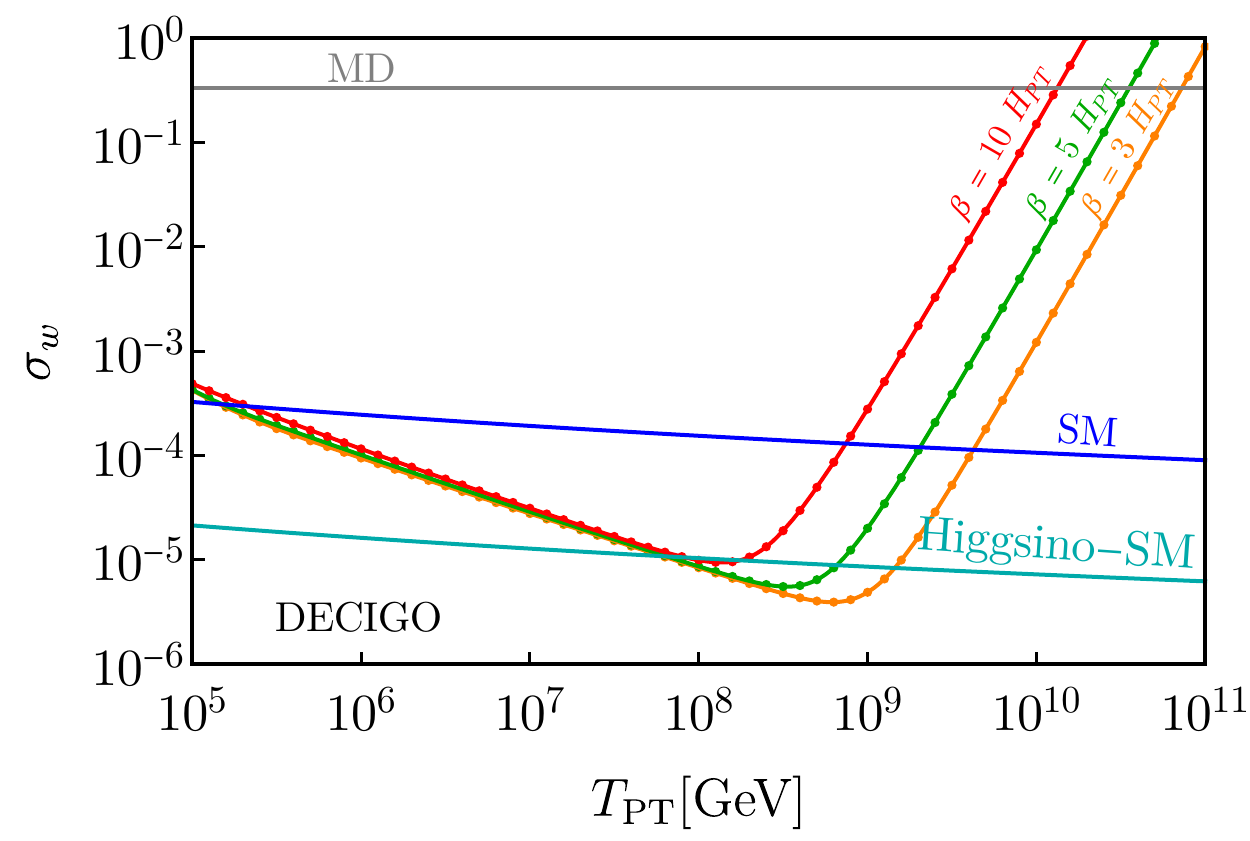}
\caption{The projected sensitivity, calculated using the Fisher matrix, of LISA (left) and DECIGO (right) to a constant equation of state near $w = 1/3$ as a function of the temperature at which the phase transition occurred for various $\beta$.  $\beta$ controls the duration of the phase transition generating the GW.  For reference we show matter domination ($\delta w = -1/3$) as well as the SM and SM plus doublet dark matter predictions for $\delta w$ versus temperature.}
\label{Fig: betaw}
\end{figure}

Next, we show how $\sigma_w$ depends on $\beta$ in Fig.~\ref{Fig: betaw}. Effectively, $\beta$ is a measurement of how sub-horizon the physics that generates the GWs is.  As such, while it has an impact on the peak of the spectrum, it does not have a large effect on the low frequency tail that we are interested in.  The main effect of $\beta$ is thus very similar to $\alpha$, where it just changes the amplitude of the low frequency tail.  As such, the qualitative behavior is quite similar.  For the DECIGO sensitivity in Fig.~\ref{Fig: betaw}, one sees that at low $T_{PT}$ where the signal is much larger than noise, $\sigma_w$ is insensitive to $\beta$, while at for high $T_{PT}$, $\sigma_w$ is larger for smaller amplitudes (larger $\beta$).

\begin{figure}[t]
\centering
\includegraphics[width=.49\linewidth]{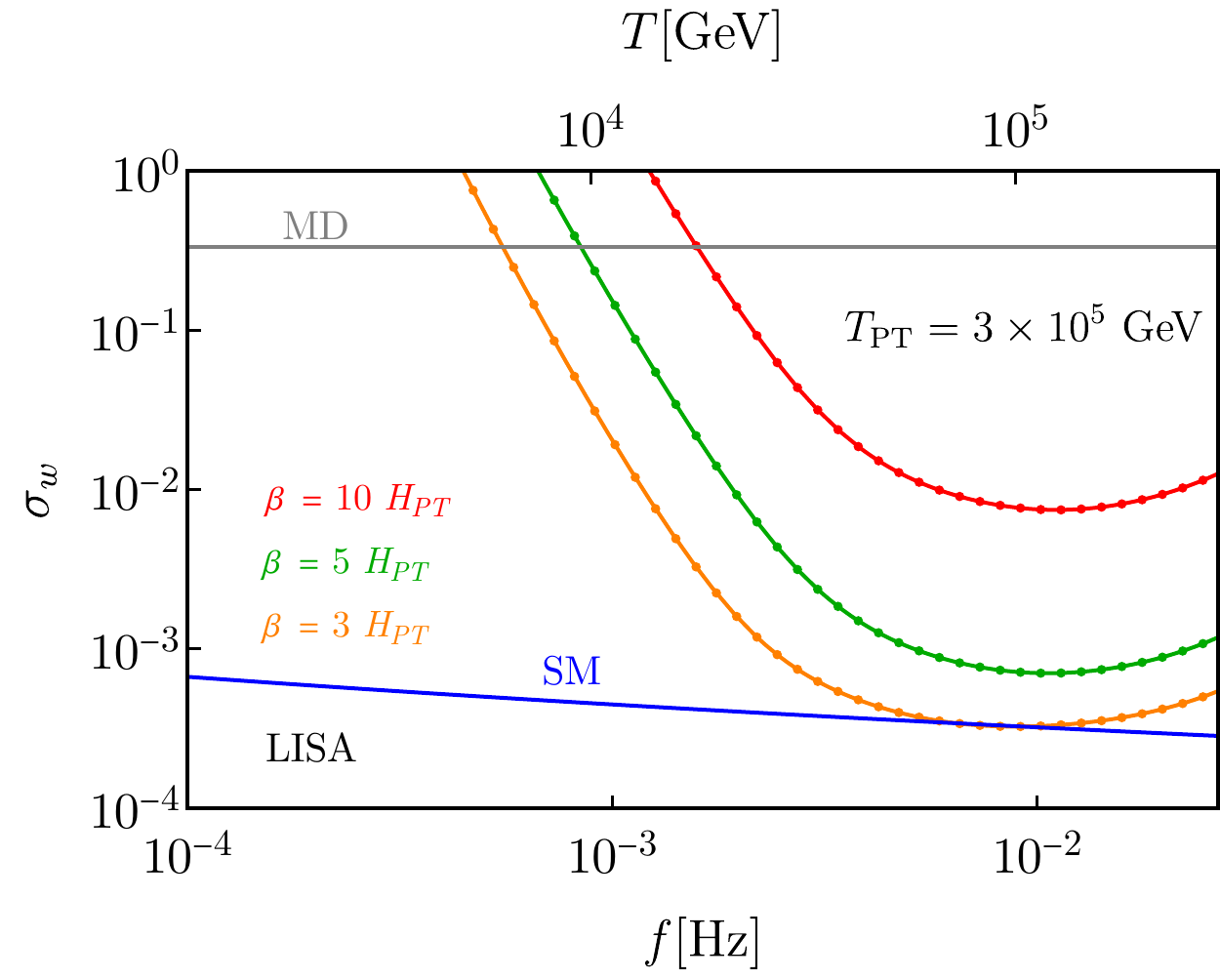}
\includegraphics[width=.49\linewidth]{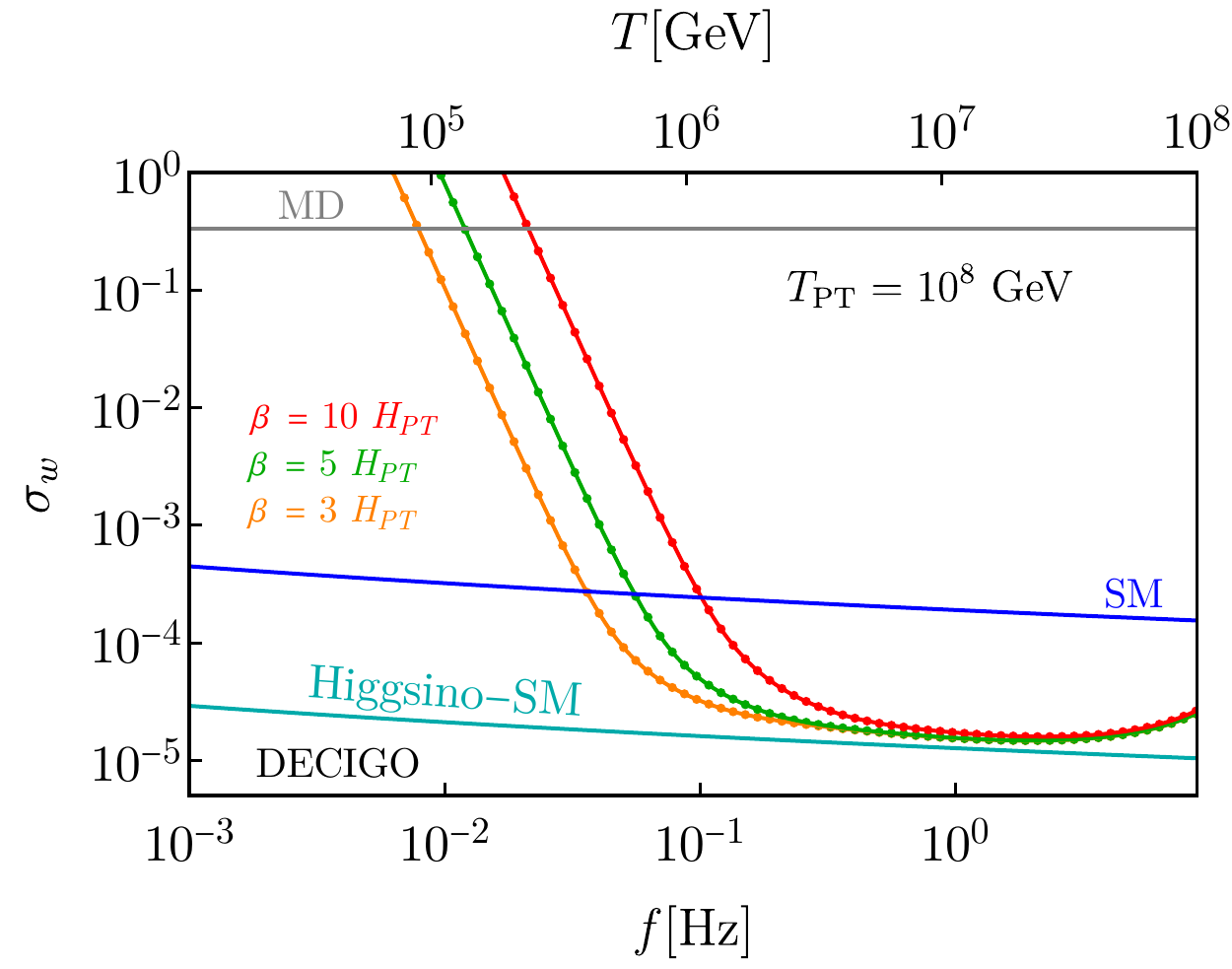}
\caption{The projected sensitivity, calculated using the Fisher matrix, of LISA (left) and DECIGO (right) to the equation of state near $w = 1/3$ as a function of frequency for various $T_{PT}$.  The sensitivity was obtained by binning the data logarithmically between a frequency $f_0$ and $f_0/3$.}
\label{Fig: w(T)}
\end{figure}

In the previous two examples, we assumed that $\delta w$ was temperature independent.  One of the most exciting prospects would be if it is possible to measure $w(T)$ as a function of temperature.  Assuming radiation domination from the temperature of $T_{PT}$ to matter-radiation equality gives a one-to-one mapping between the measured frequency dependence and the desired temperature dependence (ignoring the small correction due to $\delta w \neq 1/3$ which can be easily included). As such, we separated frequency space into bins logarithmically spaced from $f/3$ to $f$.  For each bin, we obtain the sensitivity to $\sigma_w$ using Eq.~\ref{Eq: fisher} limiting the integration to be only over the corresponding frequency bin.  The expected sensitivity for such analysis is shown in Fig.~\ref{Fig: w(T)}.  From this we see the remarkable result that in the advent of a loud GW signal generated at large temperatures, a precise measurement of $w(T)$ can be made over a large range of temperatures.

\subsection{Free-streaming fraction}

\begin{figure}[t]
\centering
\includegraphics[width=.49\linewidth]{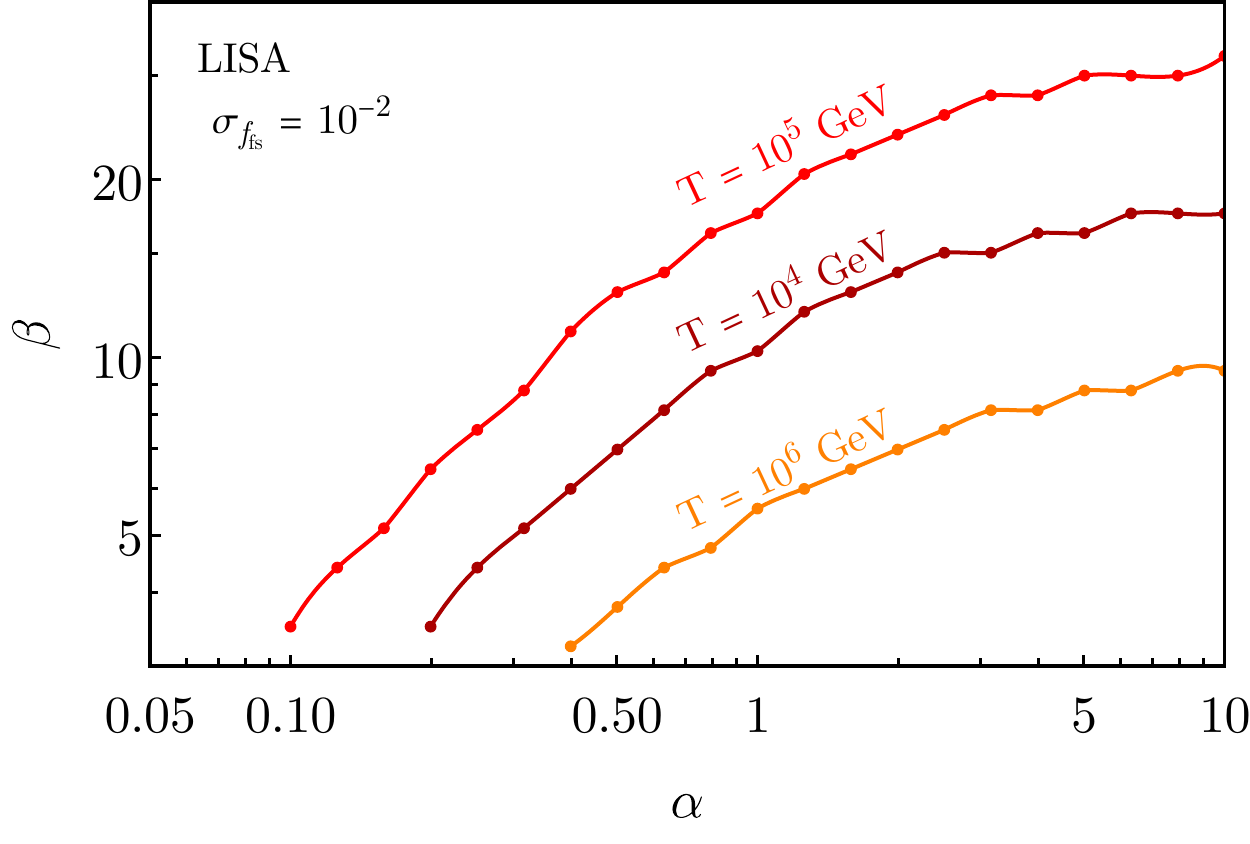}
\includegraphics[width=.49\linewidth]{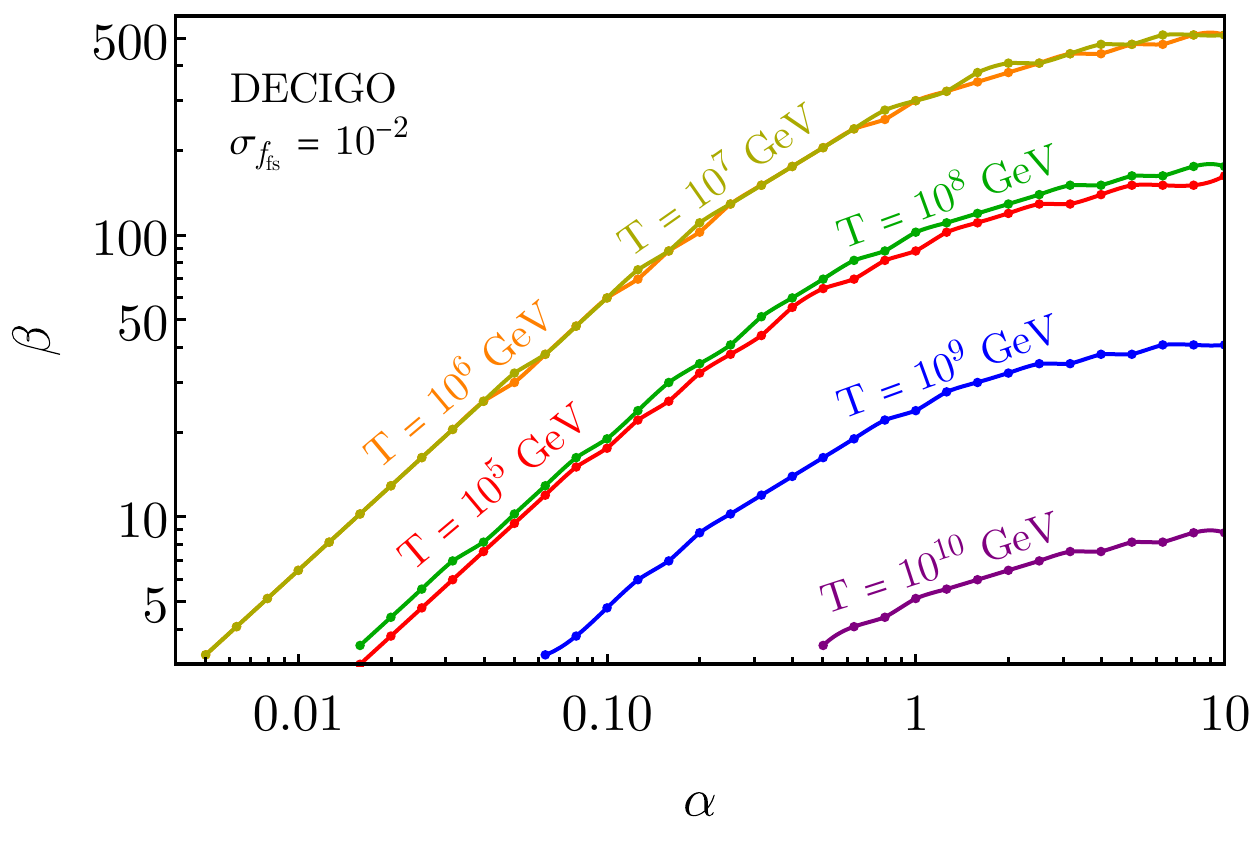}
\caption{The regions of phase transition parameter space for which the well motivated value of $f_{FS} \sim 10^{-2}$ can be reached at LISA (left) and DECIGO (right).  For a given value of $T_{PT}$, as long as the phase transition parameters $\alpha$ and $\beta$ are to the right of the line, then a sensitivity of $f_{FS} \sim 10^{-2}$ can be obtained.}
\label{Fig: alphabeta}
\end{figure}

The small free-streaming fraction limit and the small $\delta w$ limit are similar in that, in both cases, the main observable effect is the change in the slope away from the $k^3$ scaling.  As such, the results of the previous sub-section can be applied to non-zero $f_{FS}$ by using the substitution $\sigma_w \rightarrow 16 \, \sigma_{f_{FS}}/15$ to obtain the sensitivity to free-streaming particles.
If the free-streaming fraction $f_{FS} \lesssim 10^{-3}$, then at that point there is necessarily ``background" coming from the SM predicting $\delta w \sim 10^{-3}$.
In the small $\delta w$ and $f_{FS}$ limit, the similarity of these two signals leads to the unfortunate degeneracy  that GW detectors are only sensitive to the linear combination $\delta w + 16 \, \sigma_{f_{FS}}/15$.

As we will see in Sec.~\ref{Sec: application}, there is a well motivated benchmark of $f_{FS} \sim 10^{-2}$.  Given this a well defined benchmark target, we will characterize for what values of the phase transition parameters this benchmark sensitivity can be achieved assuming that $\delta \omega = 0$. In Fig.~\ref{Fig: alphabeta}, we show contours of fixed phase transition temperature $T_{PT}$ versus the phase transition parameters $\alpha$ and $\beta$.  For a given $T_{PT}$, if the phase transition parameters are to the right of the line, then a sensitivity of $f_{FS} \sim 10^{-2}$ can be achieved.

Aside from the small free streaming fraction limit, there is a unique behavior of free-streaming particles that manifests itself for large free streaming fractions.  When the free streaming fraction is large, $f_{FS} > 5/32$, it induces oscillations in the GW spectrum. The precise form of the transfer function has to be computed numerically by solving an integro-differential equation (see Sec.~\ref{Sec: free}), but for small frequencies it is well approximated by~\cite{Hook:2020phx}
\bea \label{eq:large-ffs}
F(k, f_{FS} > 5/32) \propto k^{-1} \left[1 + C \sin \left (   \log \l k \tau_{PT} \r \sqrt{\frac{32}{5} f_{FS} -1}  + \delta \right ) \right] \, ,
\eea
where $C$ and $\delta$ are $k$ independent functions of $f_{FS}$.

\begin{figure}[t]
\centering
\includegraphics[width=.7\linewidth]{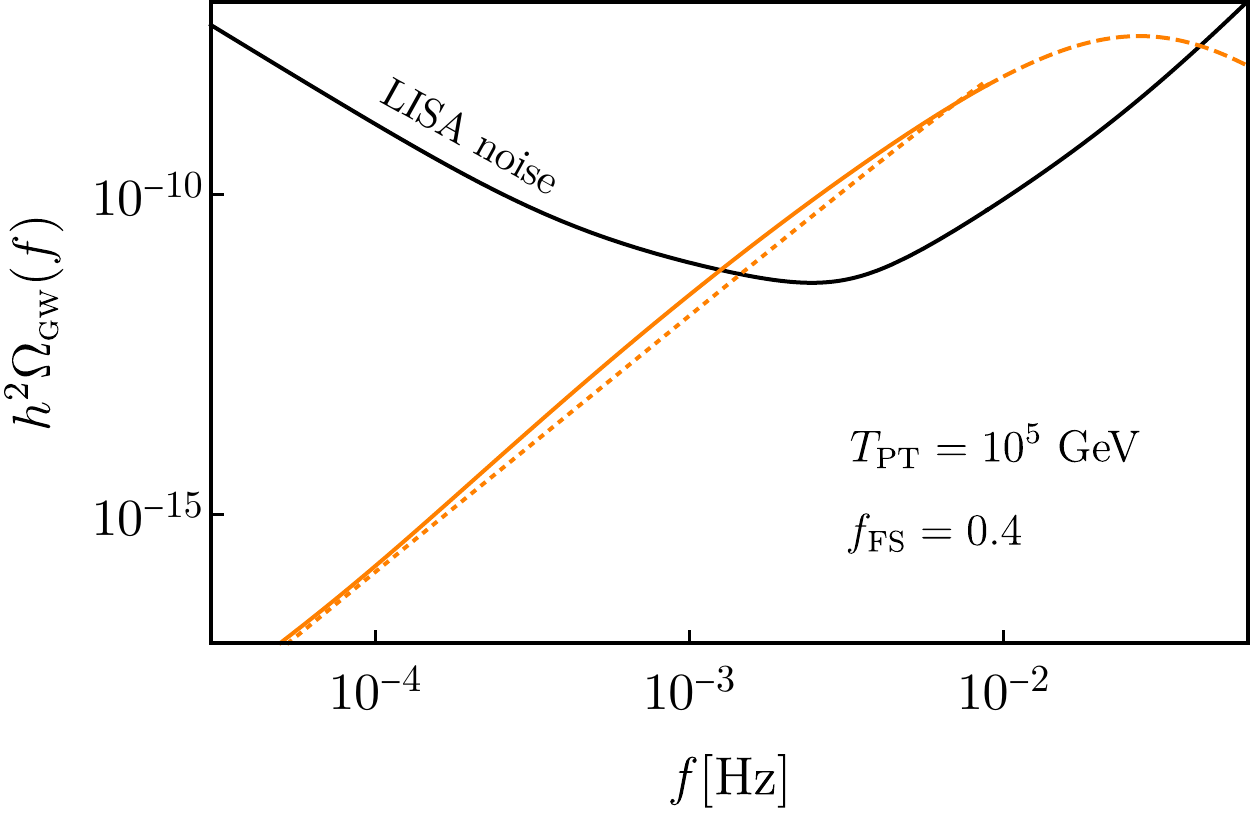}
\caption{$\Omega_{GW}(f)$ versus frequency for a free-streaming fraction of $f_{FS} = 0.4$.  The dashed line indicates the sub-horizon modes, the solid line are the super-horizon modes, and the dotted line is the $f^4$ fall off on top of which oscillations occur.  The small oscillations are what allows one to distinguish between various large values of $f_{FS}$.  For this particular example, a sensitivity of $\sigma_{f_{FS}} = 1.1 \times 10^{-3}$ can be obtained.}
\label{Fig: ffs}
\end{figure}

We now determine how accurately one can measure $f_{FS}$ in the large $f_{FS}$ limit.  The signal of large $f_{FS}$ comes in the form of oscillations on top of a $k^{-4}$ fall off.  The amplitude, phase and periodicity of the oscillation informs one about the value of $f_{FS}$.  To get an understanding of LISA's sensitivity to $f_{FS}$ in the large frequency limit, we consider the example $f_{FS} = 0.4$.  Fixing $f_{FS}$ to this value, we obtain the maximal sensitivity to $f_{FS}$ using the Fisher analysis shown in Eq.~\ref{Eq: fisher}.  
As before, we will take $\alpha = 1$, $\beta = 3$ and $v_w = 1$.
The spectrum is shown visually in Fig.~\ref{Fig: ffs}.  Using Eq.~\ref{Eq: fisher}, we find that a sensitivity of $\sigma_{f_{FS}} \sim 1.1 \times 10^{-3}$ can be reached.

\section{Implications for well motivated models}
\label{Sec: application}

When making precision measurements of $\delta w$ and $f_{FS}$, there are several benchmark points that are of great interest.  For $\delta w$, the first benchmark is at $\delta w \sim 10^{-3}$ (achievable at LISA) while the second benchmark is $\delta w \sim 10^{-5}$ (achievable at DECIGO).  For $f_{FS}$ the benchmark value is $f_{FS} \sim 10^{-2}$ (achievable at LISA).

The reason why there are any benchmarks at all for $\delta w$, is because, as reviewed in the Appendix, for the SM in a radiation dominated universe
\bea
T^\mu_\mu = \rho - 3 p \, , \qquad \delta w(T = 10^5 \, \text{GeV}) = \frac{T^\mu_\mu}{3 \rho} \approx -3 \times 10^{-4} \l \frac{100}{g_\star} \r \l \frac{\beta_{QCD}}{\beta_{QCD}^{SM}} \r.
\eea
The dependence of $\delta w$ on the total number of degrees of freedom in the universe is what makes measuring $\delta w$ so appealing.
Many motivated models predict a doubling of the SM $\beta$ functions and/or the number of degrees of freedom, resulting in a change in $\delta w$ of $\sim 10^{-3}$, giving the benchmark.  Meanwhile, if the SM is augmented by a single degree of freedom, $\delta w$ changes by around $10^{-5}$, giving the second benchmark.  A simple well motivated example which has this feature is the famous WIMP dark matter candidate.

A measurement of $f_{FS}$ to the order of $10^{-2}$ is important because that is the contribution one gets from a single new particle that freezes-out while relativistic, like the neutrinos do in the SM.  Aside from the fact that it happens in the SM, there is good reason to expect that something similar could occur in the early universe as well.  As an example, in many axion models, axions are in thermal equilibrium with the SM before freezing out.  After freezing out, they  become free-streaming particles and lead to $f_{FS}  \sim 10^{-2}$.

\begin{figure}[t]
\centering
\includegraphics[width=.7\linewidth]{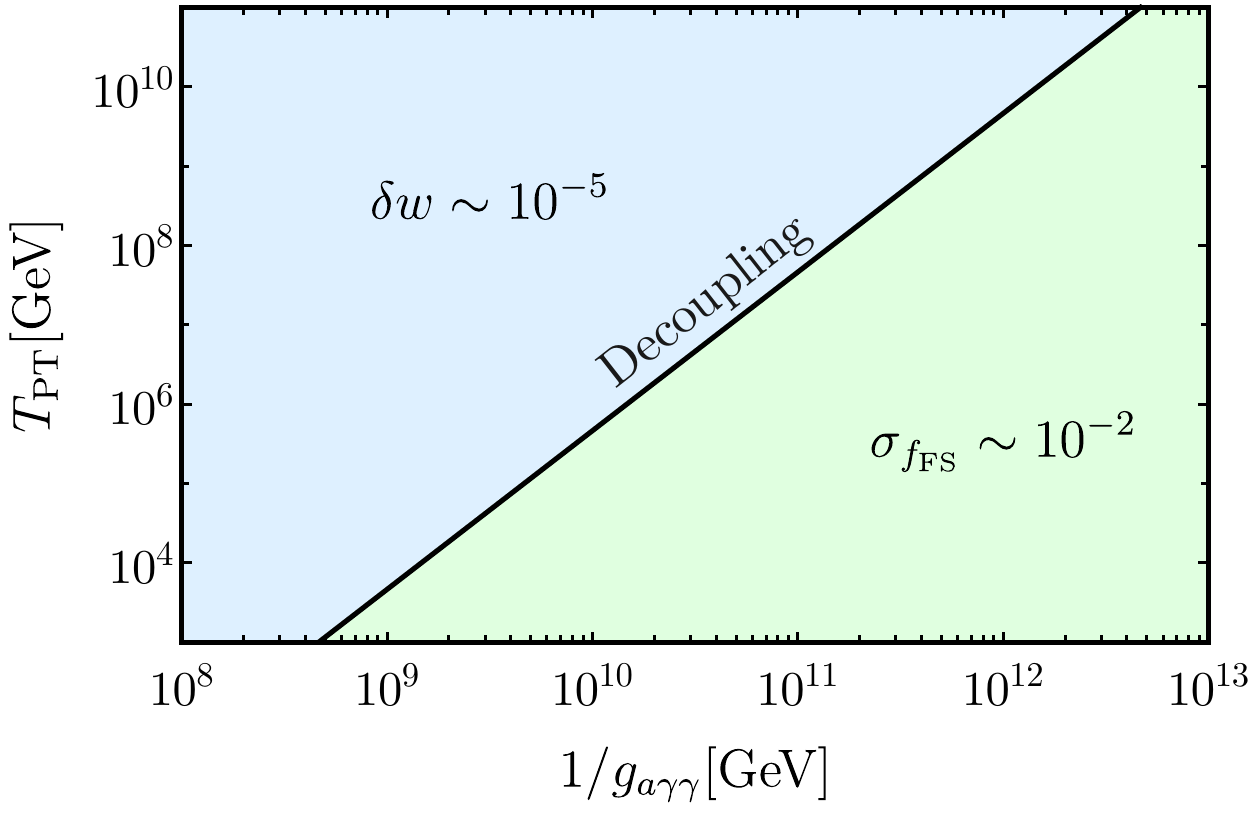}
\caption{The region of axion parameter space that can be probed as a deviation from the equation of state or as a free streaming fraction.  The black line,  $T_{dec}(g_{a \gamma \gamma})$, gives the decoupling temperature of the axion as a function of the coupling to photons.  In the blue region, axions in equilibrium with the SM predict $\delta w \sim 10^{-5}$ while in the green region, free streaming axions predict $f_{FS} \sim 10^{-2}$ so long as the reheating temperature, $T_{RH}$, satisfies $T_{RH} > T_{dec}(g_{a \gamma \gamma})$.}
\label{Fig: axion}
\end{figure}

As one particularly well motivated models for both $\delta w$ and $f_{FS}$ are axions, we show how different regions of axion parameter space can be probed with gravitational waves in Fig.~\ref{Fig: axion}.  
The black line is $T_{dec}(g_{a \gamma \gamma})$ and gives the decoupling temperature of the axions from the thermal bath as a function of the axion coupling to photons (assuming this is the largest coupling).
The region of parameter space in blue predicts $\delta w \sim 10^{-5}$.  Meanwhile, the region of parameter space in green predicts $f_{FS} \sim 10^{-2}$ as long at the reheating temperature obeys $T_{RH} > T_{dec}(g_{a \gamma \gamma})$.

Finally it is worth noting, that a non-zero measurement of $\delta w$ or $f_{FS}$ can very easily prevent a clean measurement of the other.  So while it would be extremely exciting to measure one or the other non-zero, it is unlikely that everything mentioned in this section can be realized simultaneously.

\subsection{$\sigma_w \lesssim 10^{-3}$}

The strongest motivation for TeV scale physics lies within the electroweak hierarchy problem.  Many solutions to this problem involve a large number of degrees of freedom.  Perhaps the most famous of these solutions is supersymmetry (SUSY), for a review see Ref.~\cite{Martin:1997ns}.  Supersymmetry doubles the number of degrees of freedom and predicts
\bea
\delta w (T = 10^5 \, \text{GeV}) = - 4 \times 10^{-4}.
\eea 

From Fig.~\ref{Fig: alphaw}, one can see that it is possible that LISA (DECIGO) would be able to measure $\delta w(T)$ at this accuracy for phase-transition temperatures between $\sim 10^{4-6}$ GeV ($\sim 10^{4-10}$ GeV).
As a result, both low scale and high scale SUSY can be tested at GW detectors.  Because $w(T)$ is being measured at such a large temperature, as long as supersymmetry is present at temperatures below $10^{4-6}$ GeV, LISA could be able to test it.

Aside from supersymmetry, other models such as Twin Higgs~\cite{Chacko:2005pe} and Composite Higgs~\cite{Contino:2010rs} also introduce a large number of degrees of freedom and predict deviations at the $10^{-3}$ level or larger.
Meanwhile, models where the SM is UV completed into a conformal sector give a striking signature.  Conformal sectors have $T^\mu_\mu = 0$ and thus predict $\delta w = 0$, unlike the SM prediction.
As a result, we see that testing $\delta w$ at the level of $10^{-3}$ probes almost all solutions to the electroweak hierarchy problem.

Almost all solutions to the hierarchy problem point to new physics at the TeV scale, while LISA tests if they are present below the $10^6$ GeV scale.  As a result, LISA has the potential to make a conclusive statement about our understanding of naturalness.  DECIGO would be able to push these statements to the extreme and reach even the $10^{10}$ GeV scale.

\subsection{$\sigma_w \lesssim 10^{-5}$}

Another benchmark for a measurement of $\delta w$ is $10^{-5}$.  At this level, one can test even the addition of a single degree of freedom to the SM bath at high temperatures, even if it is not free-streaming.  There are many reasons to expect at least a single new particle in thermal equilibrium with the SM, a primary one being dark matter related.
One of the simplest ways of producing dark matter is the process of thermal freeze-out.  The fact that TeV scale weakly interacting particles undergoing the process of freeze-out gives the correct relic abundance is the famous ``WIMP miracle".

Dark matter models where the abundance is set through thermal freeze-out, in general require that the dark matter's mass must be below 100 TeV.  This is because if it were any heavier, then the cross section needed to reproduce the observed dark matter density would violate the unitarity bound.  As such, most dark matter model which involves thermal freeze-out would be tested by a $\sigma_w \lesssim 10^{-5}$ measurement.
The possibility of testing the most motivated mechanism of producing dark matter, rather than testing any model specifically, is what makes this limit appealing.

Another motivation for expecting at least a single new particle in thermal equilibrium is the axion.  Axions are ubiquitous in UV complete models such as string theory and can solve problems such as the strong CP problem as well as be dark matter.  For the sake of simplicity, we will focus on the axion coupling to photons, $g_{a \gamma \gamma}$.
For a large range of parameters, the axion will be in thermal equilibrium with the SM after the GWs are generated.  In this case, there will be a prediction of $\delta w \sim 10^{-5}$.  The region of axion parameter space where this prediction is realized is shown in blue in Fig.~\ref{Fig: axion}.

\subsection{$\sigma_{f_{FS}} \lesssim 10^{-2}$}

The last benchmark value to explore is when $f_{FS} \sim 10^{-2}$.  A free streaming fraction of $10^{-2}$ is present whenever there was a particle in thermal equilibrium that subsequently freezes out while still relativistic, much like what happens to the neutrinos.  Aside from the fact that this exact process happened for neutrinos, there are other reasons to expect something similar could have happened in even earlier stages of the universe.

As in the previous example, one of the best motivated models that have free streaming particles is the axion.  For large $g_{a \gamma \gamma}$, the axion was in thermal equilibrium giving $\delta w \sim 10^{-5}$.  However for small $g_{a \gamma \gamma}$ the axion will have been in thermal equilibrium at early times and then have decoupled and become a free streaming particle in the sensitivity window of LISA/DECIGO.  Thus a sensitivity to $f_{FS} \sim 10^{-2}$ would allow one to probe much of the parameter space of the axion, namely the region of parameter space shown in green in Fig.~\ref{Fig: axion}.

A final reason why a non-zero value of $f_{FS}$ is motivated is that GWs themselves are free streaming particles.  The current best bounds on GWs, independent of their frequency, comes from $\Delta N_{\rm eff}$ constraints.  A measurement of $f_{FS} \lesssim 10^{-2}$ is a sensitivity comparable to future constraints on GWs coming from CMB S4 experiments.  Thus as long as LISA/DECIGO achieve a constraint better than this, they would be the strongest bound on new free streaming particles.  
A generic value to expect for $f_{FS}$ due to GWs themselves can be seen using the total energy in GWs due to sound waves.  This energy normalized to the total energy at the time of the phase transition is given by Ref.~\cite{Hindmarsh:2017gnf}
\bea
f_{FS} \ge \frac{\Omega_{GW}}{\Omega_{\text{total}}} \approx 5 \times 10^{-4} \l \frac{100 H_{PT}}{\beta} \r \l \frac{4 \alpha^2}{(1 + \alpha )^2} \r ,
\eea
where we are taking the bubble wall velocity to be 1 and 100\% conversion of the vacuum energy into kinetic energy.  In most models, $\alpha \sim 1$ but $\beta/H_{PT} \sim 100$, though some models can obtain smaller values of $\beta$.  As a result, a reasonable value of the free streaming fraction to expect due to gravitational waves is $f_{FS} \gtrsim 10^{-3}$.
Thus it is quite reasonable to expect that GWs themselves might be loud enough to comprise an observable value of $f_{FS}$.
The example given above is simply one source of GWs.  Different GW sources have different natural expectations for the value of $f_{FS}$.

\section{Massive free streaming particles}
\label{Sec: free}

The presence of relativistic free streaming particles affects the propagation of gravitational waves, since the energy momentum tensor of such particles develops an anisotropic stress in the presence of gravitational waves~\cite{Weinberg:2003ur,Hook:2020phx}. In this section, we generalize this result by considering what happens when free streaming particles have a mass in order to study the impact of free streaming particles transitioning from being relativistic to non-relativistic.
For the CMB, the fact that neutrinos are becoming non-relativistic around the time when the CMB is being generated leads to important imprints.
For stochastic GWs, we find that as free streaming particles become non-relativistic, their damping effect on the GW spectrum vanishes.  The more e-foldings that the free streaming particles are relativistic after GW generation, the more they dampen the GWs.

\subsection{Derivation of the damping}

We start with the metric for tensor perturbations in an isotropic background, which is given by
\bea \label{eq: metric}
    g_{\mu\nu} &=& a^2(t) (\delta_{\mu\nu }+h_{\mu\nu}(\vec{x},t)),\\
    h_{0 \nu} &=& 0, \nonumber
\eea
where $a(\tau)$ is the scale factor and $h_{\mu \nu}$ is a perturbation of the metric. We make a standard choice of transverse-traceless gauge (TT), meaning that $h_{ij,j} = 0$ and $h^{i}_{i} = 0$. The evolution of the remaining components of $h_{ij}$ is governed by the linearized Einstein equation
\begin{equation} \label{eq: lin ein-bol}
    -\frac{1}{2} h_{ij;\nu}^{;\nu} = 8\pi G \Pi_{ij} \, ,
\end{equation}
where $\Pi_{ij}$ is the anisotropic part of the energy-momentum tensor $T_{ij} = p g_{ij}+a^2 \Pi_{ij}$.
Knowing the form of the metric, defined in Eq.~\eqref{eq: metric}, we can unwrap covariant derivatives leading to the well-known equation
\begin{equation} \label{eq: ein-bol fourier}
  h_{\lambda,k}''(\tau) + 2 \frac{a'}{a} h_{\lambda,k}'(\tau) + k^2 h_{\lambda,k}(\tau) = 16 \pi G a^2 \Pi_{\lambda,k}. 
\end{equation} 
Here, the metric perturbations, $h_{ij}$, and anisotropic energy-momentum tensor, $\Pi_{ij}$, have been written in the momentum $k$ and polarization $\lambda$ space as
\bea
    h_{ij}(\tau,x) &=& \sum_{\lambda = +,\times} \int \frac{d^3 k}{(2\pi)^3} h_{\lambda,k} (\tau) e^{i k x} \epsilon^{\lambda}_{ij},\\
    \Pi_{ij}(\tau,x) &=& \sum_{\lambda = +,\times} \int \frac{d^3 k}{(2\pi)^3} \Pi_{\lambda,k} (\tau) e^{i k x} \epsilon^{\lambda}_{ij}, \label{Eq: Pi fourier}
\eea
where $\epsilon^{+,\times}_{ij}$ are the polarization tensors. Our aim now is to determine the form of $\Pi_{\lambda,k}$ to linear order in the perturbations. The energy-momentum tensor is given by
\bea \label{eq: Tij def}
    T_{ij} &=& \frac{1}{\sqrt{-\det g}} \int d^{3} P \, F(\vec{x},\tau, P_i) \frac{P_{i} P_{j}}{P^{0}}
    \eea
where $F(x^i,\tau,P_i)$ is the phase space density and $P^{\mu} \equiv \frac{d x^{\mu}}{d \lambda}$. Switching to comoving momentum, $q^\mu$, defined by
\begin{equation}
	P_{i}  = \left(\delta_{ik}+\frac{h_{ik}}{2}\right)q_{k}, 
	\label{eq:comoving-momentum}
\end{equation}
with $q_i = q \gamma_i$, where $\gamma_i$ are the directional cosines, and doing a perturbative expansion for the phase-space distribution $F(\vec x,\tau,q,\vec \gamma) = F_0(\tau, q) + F_1(\vec x,\tau,q,\vec \gamma) + \dots$, one finds at first order
\begin{equation} \label{eq: Pi}
\Pi_{ij} = a^{-4}(\tau) \int d^{3} q \, F_1(\vec{x},\tau,q,\gamma_i) \frac{q^2 \gamma_i \gamma_j}{\sqrt{q^2+a^2 m^2}}.
\end{equation}

For simplicity, we will assume that free streaming particles were once in thermal equilibrium and at some point decoupled from the rest of the bath.  While this assumption is not necessary, it is the case for many motivated examples of free streaming particles.
When a given particle species is in thermal equilibrium, its 0th order phase-space density is given by the thermal distribution in a homogeneous background
\begin{equation}
F(\vec{x},\tau,q,\vec{\gamma}) =  \frac{N}{(2\pi)^3} \left[ \exp{\l \frac{\sqrt{q^{2}+m^2 a(\tau)^2}}{a(\tau) k T}\r}  \pm 1 \right]^{-1}  ,
\end{equation}
where $N$ is the number of degrees of freedom in that species.
Assuming that a given species $\chi$ decouples at $\tau=\tau_{d}$, the evolution of the phase space density freezes 
\begin{equation} \label{eq: F0 decoupling}
F_0(\tau>\tau_d,q)  =  \frac{N}{(2\pi)^3} \left[ \exp{\l \frac{\sqrt{q^{2}+m^2 a(\tau_d)^2}}{a(\tau_d) k T_d}\r}  \pm 1 \right]^{-1} \approx \frac{N}{(2\pi)^3} \left[ \exp{\l \frac{q}{a(\tau_d) k T_d}\r}  \pm 1 \right]^{-1}
\end{equation}
where we assume that $\chi$ decouples when $T_d \gg m$.  

We can compute the evolution of the perturbation $F_{1}(\vec{x},\tau,q^{0},\vec{\gamma})$ by expanding the Boltzmann equation
\begin{equation}
\frac{d F}{d \tau} = \frac{\d F}{\d \tau} + \frac{d x^{i}}{d \tau} \frac{\d F}{\d x^{i}} + \frac{d q}{d \tau} \frac{\d F}{ \d q} + \frac{d \gamma^{i}}{d \tau} \frac{\d F}{ \d \gamma^{i}} = 0 \, .
\end{equation}
To leading order in $h$, the Boltzmann equation simplifies to
\begin{equation}
\frac{\d F_1}{\d \tau} + \gamma_{i} v  \frac{\d F_1}{\d x^{i}} = \frac{1}{2} h_{ij}' \gamma_{i} \gamma_{j} q\frac{\d F_0}{ \d q},
\end{equation}
where $v = q/\sqrt{q^2 + a^2 m^2}$. Going to Fourier space and using the polarization vectors we can decompose the perturbation $F_1$ into
\begin{equation}
	F_1(\vec x, \tau, q, \vec \gamma) = \sum_{\lambda = +,\times} \int \frac{d^3 k}{(2\pi)^3} e^{i \vec k \cdot \vec x} f_k(\tau, \mu, q)  \epsilon^{\lambda}_{ij} \gamma_i \gamma_j ,
\end{equation}
where $\mu = \hat k \cdot \vec \gamma$, leading to
\begin{equation} \label{eq:F1-fourier}
\frac{\d f_{k,\lambda}}{\d \tau} + i v k \mu  f_{k,\lambda} =  q \frac{\d F_0}{\d q}\frac{1}{2} \frac{\d h_{\lambda,k} }{\d \tau} \, .
\end{equation}
Assuming that there were no tensor perturbations in the free streaming radiation before the gravitational waves were generated, we have
\begin{equation} \label{eq: f_k}
	f_{k,\lambda}(\tau,q,\mu) = \frac{1}{2} q \frac{\d F_{0}}{\d q} \int\limits_{\tau_{PT}}^{\tau} d \tau' h'(\tau') e^{-i \int\limits_{\tau'}^{\tau} d \tau'' v(\tau'') k \mu  } \, .
\end{equation}
Plugging the above result into Eq.~\ref{eq: Pi}, one finds after some algebra
\begin{equation} \label{eq: Pi final}
\Pi_{\lambda,k} = a^{-4} \int d q \frac{q^5}{\sqrt{q^2+ a^2 m^2}} \frac{\d F_{0}}{\d q} \int\limits_{\tau_{d}}^{\tau} d \tau' h' (\tau ') 4 \pi  \frac{j_{2} (u_{q}(\tau')-u_{q}(\tau))}{(u_{q}(\tau')-u_{q}(\tau))^2},
\end{equation}
where $j_2$ is the spherical Bessel function, and with $u_q$ defined by
\begin{equation}
u_{q}(\tau) = k \int_{\tau_{PT}}^\tau d \tau' v(\tau') .
\end{equation}

Combining results from Eq.~\eqref{eq: ein-bol fourier} and Eq.~\eqref{eq: Pi final} we get the general equation for how massive free streaming particles affect gravitational waves:
\begin{equation} \label{eq: ein-bol final}
\begin{aligned}
h_{\lambda,k}''(\tau) + 2 \frac{a'}{a} h_{\lambda,k}'(\tau) +  k^2 h_{\lambda,k}(\tau) =  \frac{ 24 \pi f_{FS}(\tau) }{\rho_{FS}(\tau)} & \l \frac{a'}{a} \r^{2}  \int \frac{d q}{a^4} \frac{q^5}{\sqrt{q^2+ a^2 m^2}} \frac{\d F_{0}}{\d q}  \times \\
&  \int\limits_{\tau_{d}}^{\tau} d \tau' h' (\tau ')   \frac{j_{2} (u_{q}(\tau')-u_{q}(\tau))}{(u_{q}(\tau')-u_{q}(\tau))^2} \, ,
\end{aligned}
\end{equation} 
where $\rho_{FS}$ is the average energy density in free streaming particles and $f_{FS}(\tau)$ is the free streaming fraction,~Eq.~\ref{eq:ffs-definition}. Note that in the massless limit, $u_q(\tau) = k \tau$, from which follows that in a radiation dominated universe Eq.~\eqref{eq: ein-bol final} reduces to the previously known expression~\cite{Weinberg:2003ur,Hook:2020phx}
\begin{equation} 
h_{\lambda,k}''(\tau) + 2 \frac{a'}{a} h_{\lambda,k}'(\tau) + k^2 h_{\lambda,k}(\tau) = - 24  f_{FS}  \l \frac{a'}{a} \r^{2} \int\limits_{\tau_{d}}^{\tau} d \tau' h' (\tau ') \frac{j_{2} (k(\tau'-\tau))}{(k(\tau'-\tau))^2} \, .
\end{equation} 

\subsection{Numerical results}

\begin{figure}[t]
    \centering
    \includegraphics[width=.7\linewidth]{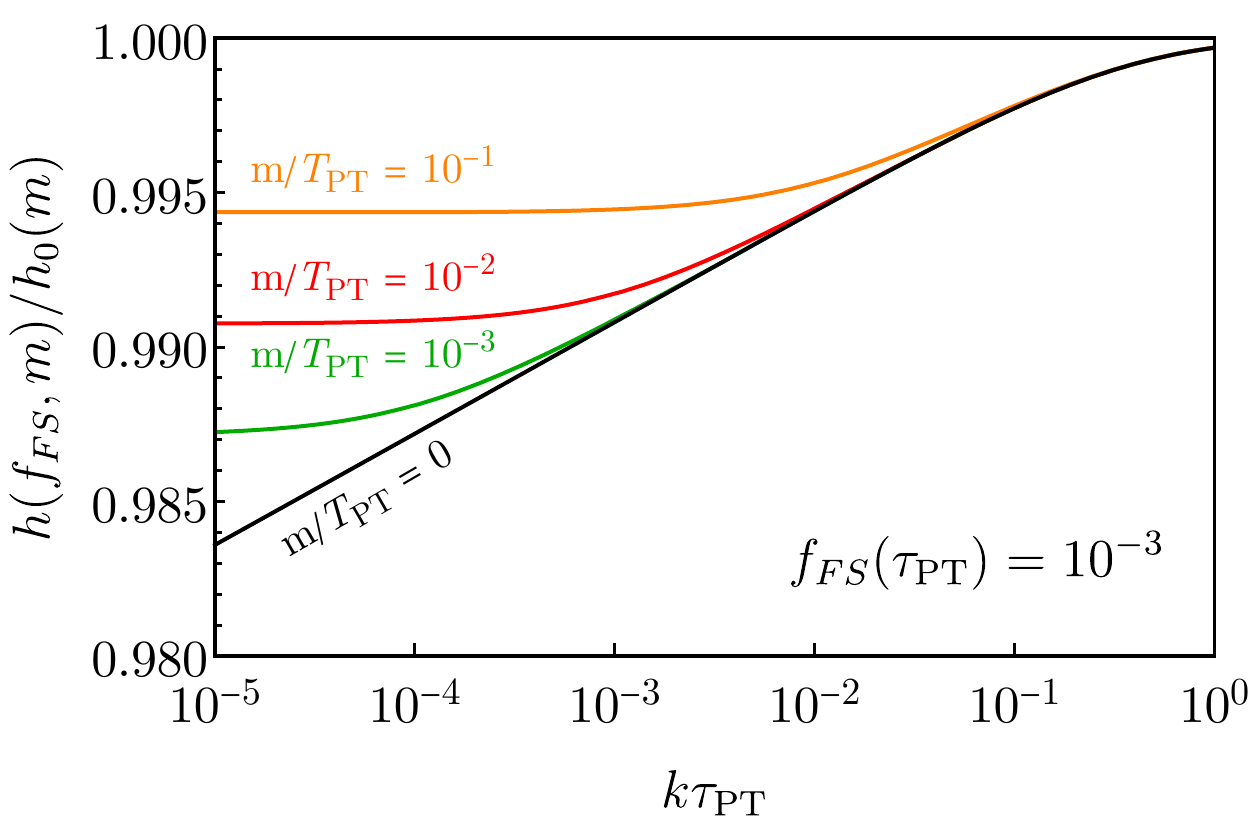}
    \caption{The fractional suppression of GWs due to massive free streaming particles as compared to massive interacting particles.  We fix the free streaming fraction to be $f_{FS}(\tau_{PT}) = 10^{-3}$ and let the mass vary $m/T_{PT} = 10^{-1}, 10^{-2}, 10^{-3}, 0$, shown in orange, red, green and black respectively.  When the free streaming particles becomes non-relativistic, they cease to suppress the gravitational waves leaving only the suppression that resulted from back when they were relativistic.}
    \label{fig: massive free streaming}
\end{figure}

In this subsection, we present numerical solutions to the integro-differential equation~\ref{eq: ein-bol final} for different values of the free streaming particle's mass. 
Massive free streaming particles modify the gravitational wave spectrum via two effects, by changing the equation of state of the universe when they become non-relativisitic, and by causing a suppression due to their free streaming nature.
When the free streaming particles transition from being relativistic to non-relativistic, the change in the equation of state can drastically affect the expansion history (e.g., if these free streaming particles where stable and didn't annihilate they could lead to a period of early matter domination). 

In order to isolate the effects on the GW spectrum due to particles free streaming from the ones coming from changes in the expansion history, we will compare ratios of the GW spectrum between two scenarios with the same expansion history.
We include a new species that starts off relativistic and later transitions to the non-relativistic regime.  In the first case this component is free streaming, $h(f_{FS},m)$, while in the second case it is not, $h_0(m)$.
In Fig.~\ref{fig: massive free streaming}, we show the suppression $h(f_{FS},m)/h_0(m)$ taking $f_{FS}(\tau_{PT}) = 10^{-3}$ for several masses $m/T_{PT} = 10^{-1}, 10^{-2}, 10^{-3}, 0$ in orange, red, green and black respectively.  

In the massless case, the suppression is larger for smaller frequencies, causing a change in the shape of the GW spectrum as explored in the previous sections.  
For the massive case, at high frequencies that enter the horizon while the free streaming particles are relativistic, the suppression tracks the massless case.  For low frequencies that enter the horizon after the free streaming particles are non-relativistic, the suppression asymptotes to a constant value at low frequencies.  For the low frequency part of the spectrum, the majority of the suppression is from when the modes were super horizon and slow rolling in the potential generated by the relativistic free streaming particles, as discussed in Ref.~\cite{Hook:2020phx}.

\section{Conclusion} \label{Sec: conclusion}

In this article, we have demonstrated that LISA (DECIGO) can potentially measure the equation of state of the universe ($w$) and/or the fraction of free streaming particles ($f_{FS}$) down to an accuracy of $10^{-4}$ ($10^{-6}$).   This measurement is analogous to 21-cm measurements in that a deformation to a known frequency distribution is used to infer propagation effects.
To illustrate the physics potential of such a precise measurement, we presented several benchmark models that predict deviations from an equation of state of $1/3$ at varying levels.  The Standard Model itself predicts a $w-1/3 \sim 10^{-3}$ deviation, well motivated solutions to the electroweak hierarchy problem predict different $w-1/3 \gtrsim 10^{-3}$ deviations and dark matter models predict $w-1/3 \sim 10^{-5}$.  Many of these same models also predict a large $f_{FS} \sim 10^{-2}$.  If LISA were to see stochastic GWs and were able to make this precision measurement, it would revolutionize our understanding of early universe physics.

The possibility of reaching these exciting benchmark values at LISA pushes for an understanding of all signals and noises at the $10^{-4}$ level, including contamination from astrophysical foregrounds, which would likely be significant for mid-band detectors like DECIGO.  In this article, we assumed that the GW signal was generated in a short sub-horizon timescale and thus the shape of the frequency spectrum is fixed, e.g. sounds waves can be sub-horizon~\cite{Ellis:2019oqb} or super-horizon~\cite{Hindmarsh:2015qta}.  It is possible that some of the source of GWs may persist for longer than a Hubble time and thus contaminate the precise measurement of the equation of state.  Our results thus motivate an improved understanding of all sources of GWs and their low frequency behavior.  
For example in a radiation dominated universe, $\Omega_{GW} \propto k^3 P_{GW}(k)$ with $P_{GW}(k)$ being the Fourier transform of the two point function of the source of GWs.  For a causal source, there exists a radius R such that $P_{GW}(x > R) = 0$ implying that $P_{GW}(k \ll 1/R) \sim c_0 + c_1 k^2 R^2$ for some constants $c_0$ and $c_1$.  There is thus necessarily a model dependent $k^5$ correction to the $k^3$ scaling that needs to be accounted for.

If LISA were to discover stochastic GWs, GW physics could instantly become a precision science.  It is exciting to see that in this case, that LISA would be able to teach us about the early universe to an unprecedented precision.

\section*{Acknowledgement}
The authors were supported in part by the NSF grants PHY-1914480, PHY-1914731, by the Maryland Center for Fundamental Physics (MCFP). GMT was also partly funded by the US-Israeli BSF Grant 201823. GMT thanks the Aspen Center where part of this work was completed, which is supported by National Science Foundation grant PHY-1607611.  The participation of GMT at the Aspen Center for Physics was supported by the Simons Foundation.


\appendix

\section{Calculation of $w(T)$}

In this section, we show how one calculates $\delta w(T) = w(T) - 1/3$ in various theories.  For the sake of simplicity, we will work in the limit where all particles are massless.
Our starting point is the first law of thermodynamics $dU = T dS - p dV$.  We can express the total energy (entropy) in terms of the energy (entropy) density using $U = \rho(T) V$ ($S = s(T) V$).  Equating the $dT$ and $dV$ terms on both sides of the first law of thermodynamics, we arrive at
\bea
s = \frac{\rho +p}{T} \qquad \rho = T \frac{dp}{dT} - p.
\eea
Pressure can be exchanged with the free energy density using $f = \rho - T s = - p$.
After a bit of algebra, we find
\bea \label{Eq: 3w}
\delta w(T) = \frac{T^4}{3 \rho} \frac{d ( f/T^4 )}{d \log T}.
\eea
Expressing the free energy as a Taylor series in terms of coupling constants, we see that the right hand side is proportional to beta functions and inversely proportional to $g_\star$.
Connected vacuum diagrams give the log of the partition function, and hence up to a minus sign give the free energy density ( $F = - T \log Z$).  The leading contribution to $\delta w$ can thus be found by calculating all connected 2-loop vacuum diagrams and taking the appropriate derivatives.

\begin{figure}[t]
    \centering
    \includegraphics{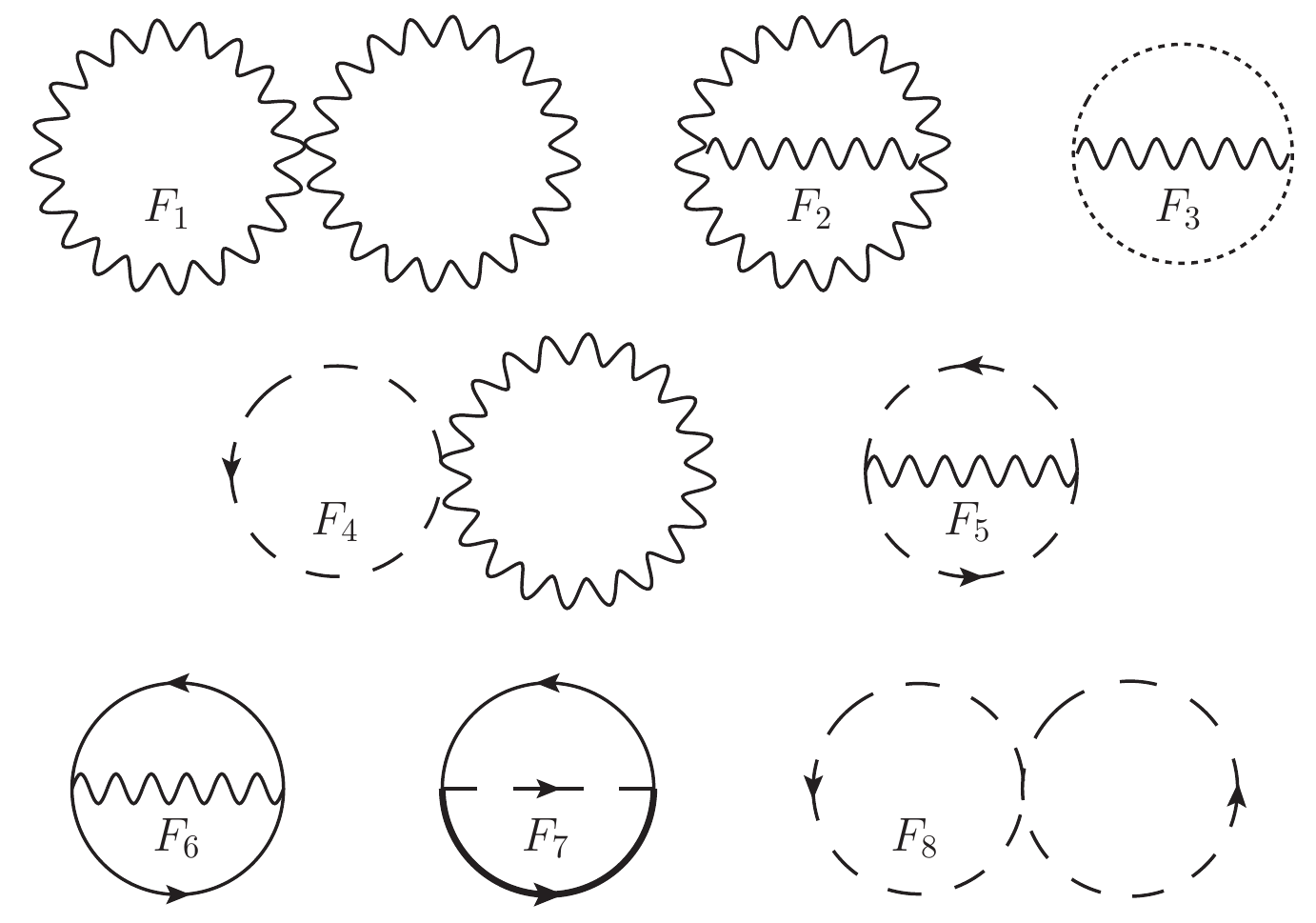}
    \caption{The eight leading order diagrams contributing to the free energy that lead to $w(T) \ne 1/3$.  $F_{1,2,3}$ involve gauge bosons and/or ghosts and are present for any non-abelian gauge theory.   $F_{4,5}$ ($F_{6}$) are present whenever there is a scalar (fermion) charged under a gauge group.  $F_7$ is the leading diagram for Yukawa couplings while $F_8$ is the leading diagram for quartic couplings.}
    \label{fig: diagrams}
\end{figure}

The 2-loop vacuum diagrams that we are interested in are shown in Fig.~\ref{fig: diagrams}.  There are eight diagrams that can all be calculated and their high temperature/low mass results are most easily expressed in terms of
\bea
I = \frac{T^2}{12}\qquad \Tilde{I} = -\frac{T^2}{24}.
\eea
The first three diagrams are present in any non-Abelian gauge theory and evaluate to
\bea
F_1 = 3 g^2 C_2(G) d(G) I^2 \qquad 
F_2 = - \frac{9}{4} g^2 C_2(G) d(G) I^2 \qquad
F_3 = \frac{1}{4} g^2 C_2(G) d(G) I^2,
\eea
where $d(G)$ is the dimension of the group and $C_2(G)$ is its quadratic casmir.  The next two diagrams are present in any gauge theory with charged scalars
\bea
F_4 = 4 g^2 C_2(R) d(R) I^2 \qquad
F_5 =  - \frac{3}{2} g^2 C_2(R) d(R) I^2,
\eea
where as before $d(R)$ ($C_2(R)$) is the dimension (quadratic casmir) of the representation.  For abelian theories $C_2(R) = Q^2$.  Diagram $F_{6}$ is present for theories with charged Weyl fermions and evaluates to
\bea
F_6 = g^2 C_2(R) d(R) \left  ( \Tilde{I}^2-2 \Tilde{I} I \right).
\eea
If there are Yukawa couplings involving a complex scalar $\Phi$ and two Weyl fermions $\psi$ and $\psi^c$ of the form $\mathcal{L} = y \Phi \psi \psi^c$, diagram $F_7$ will be present
\bea
F_7 = y^2  \left  ( \Tilde{I}^2-2 \Tilde{I} I \right).
\eea
Finally if there are scalar quartic couplings, then $F_8$ will be present.  For a quartic coupling of the form, $\mathcal{L} = \lambda (H_i H_i^\dagger)^2$, we have
\bea
F_8 = \lambda N_H (1 + N_H) I^2
\eea
where $N_H$ is the number of scalars that the index $i$ runs over.  For a different quartic coupling, e.g. those present in SUSY, the combinatorics factor, $N_H (1 + N_H)$, will be different.

\begin{figure}[t]
    \centering
\includegraphics[width=.7\linewidth]{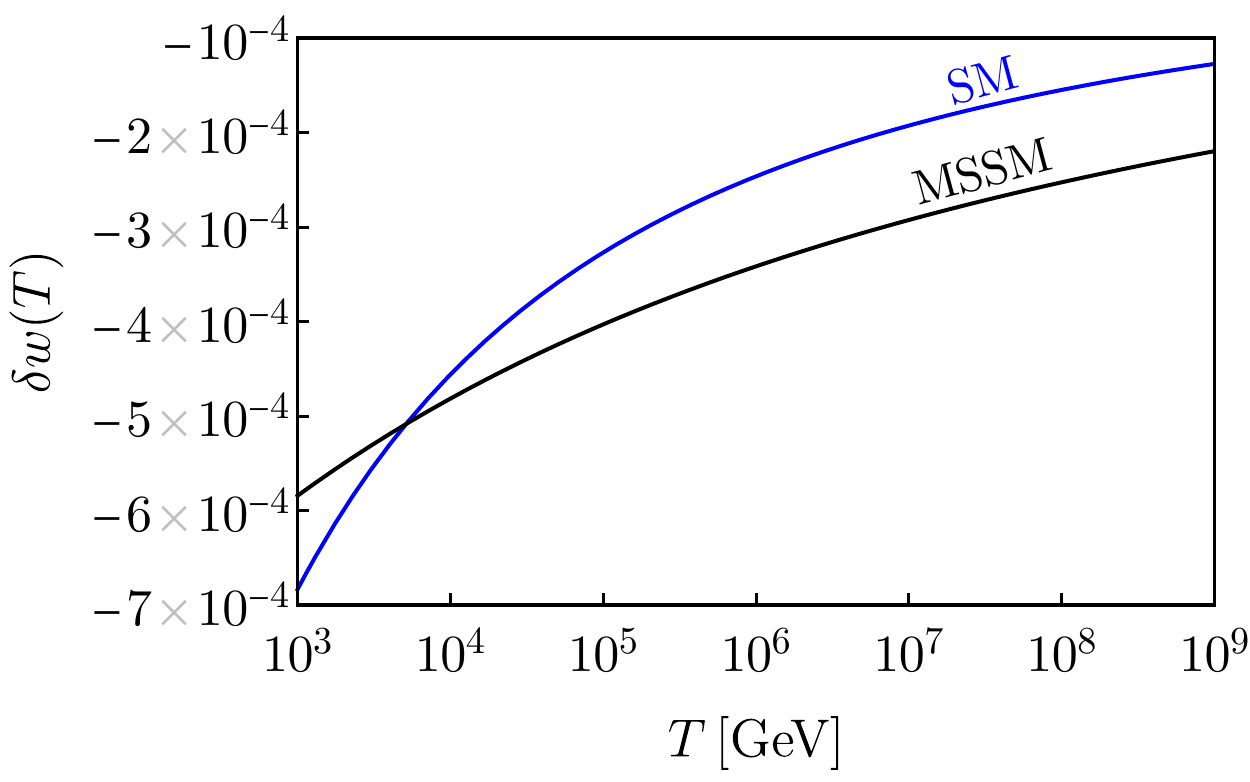}
    \caption{The deviation of equation of state from $1/3$ of the Standard Model (black) and the MSSM (blue) as a function of temperature.}
    \label{fig: SMw}
\end{figure}

We can now use the previous results to find the free energy density ($f = \sum F_i$) and combine it with Eq.~\ref{Eq: 3w} and beta functions to find $\delta w$ for various theories and at various temperatures.

\paragraph{The Standard Model} : 
When calculating $\delta w$ in the SM, we only take $y_t, \lambda, g, g'$ and $g_s$ to be non-zero and ignore all other couplings.  The dominant contribution to $\delta w$ comes from the QCD beta function because of both the abundance of colored particles and the size of the beta function.  The top Yukawa beta function is the next most important contribution and contributes only about 5\% of the final result.   
More explicitly, we find that at leading order
\bea
\delta w = \frac{T^4}{\rho} \left ( \frac{55}{1728} \beta_{g'^2} + \frac{43}{576} \beta_{g^2} + \frac{7}{36} \beta_{g_s^2} + \frac{5}{288} \beta_{y_t^2} + \frac{1}{72} \beta_{\lambda}    \right ),
\eea
where $\beta$ are the beta functions defined as $\beta_\lambda = d\lambda/d\log \mu$.
The value of $\delta w(T)$ for the SM is shown as the black line in Fig.~\ref{fig: SMw}.

\paragraph{The Minimal Supersymmetric Standard Model} :
The MSSM contains many more particles and interactions, but its QCD beta function is smaller than the SM value.  As a result of these effects partially canceling, the MSSM value of $\delta w$ is not very different from the SM value. 

Because the Higgs quartic coupling is determined by gauge couplings, we only consider contributions from $y_t, g, g'$ and $g_s$.  As with the SM,  the dominant contribution again comes from the QCD beta function with all other couplings playing an even smaller role than before.  
We find that at leading order
\bea
\delta w = \frac{T^4}{\rho} \left ( \frac{33}{64} \beta_{g'^2} + \frac{69}{64} \beta_{g^2} + \frac{21}{8} \beta_{g_s^2} + \frac{9}{32} \beta_{y_t^2}   \right ).
\eea
The value of $\delta w(T)$ for the MSSM is shown as the blue line in Fig.~\ref{fig: SMw}.

\begin{figure}[t]
    \centering
\includegraphics[width=.7\linewidth]{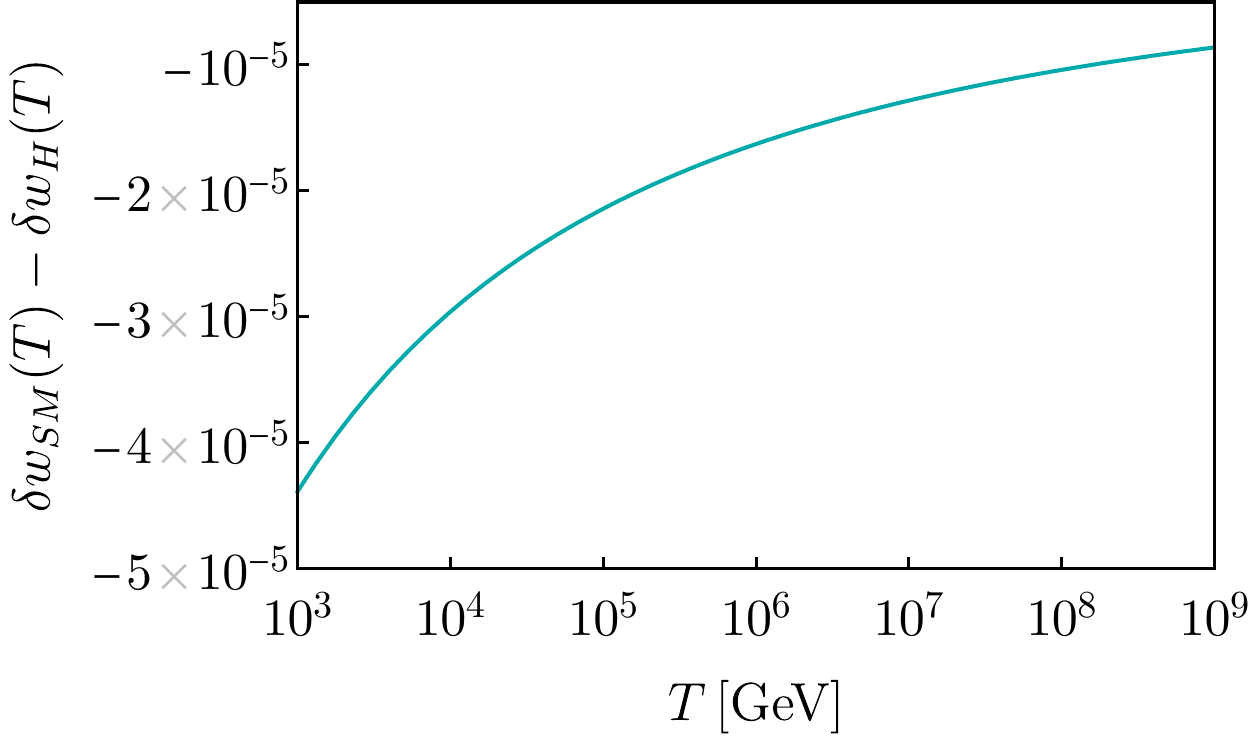}
    \caption{The difference in the equation of state of the Standard Model with and without a vector-like doublet dark matter.}
    \label{fig: diff}
\end{figure}

\paragraph{The Standard Model with doublet dark matter} :
The next model we consider is dark matter as a vector-like fermion with the quantum numbers of the Higgs boson.  This is a particularly appealing version of WIMP dark matter as it makes the SM gauge couplings unify better~\cite{Mahbubani:2005pt}.  Direct detection constraints imply that this WIMP necessarily mixes with an additional singlet, but this singlet can be much heavier than the WIMP and thus we will neglect it.  
As before, we only consider the couplings $y_t, \lambda, g, g'$ and $g_s$.  As the doublet is not color charged, it does not change the result by much.  
We find that
\bea
\delta w = \frac{T^4}{\rho} \left ( \frac{5}{144} \beta_{g'^2} + \frac{1}{12} \beta_{g^2} + \frac{7}{36} \beta_{g_s^2} + \frac{5}{288} \beta_{y_t^2} + \frac{1}{72} \beta_{\lambda}    \right ).
\eea
The difference between this result and the SM is shown in Fig.~\ref{fig: diff}.

\bibliographystyle{JHEP}
\bibliography{gw-references}

\end{document}